\documentclass[amsmath,amssymb,superscriptaddress,nobalancelastpage,prb,twocolumn,longbibliography]{revtex4-1}

\usepackage{hyperref}
\usepackage{graphicx}
\usepackage{varioref}
\usepackage{xr-hyper}
\usepackage{xcolor}
\usepackage{nicefrac}
\usepackage{xfrac}
\usepackage{hyperref}
\hypersetup{colorlinks,linkcolor=blue,urlcolor=blue,citecolor=blue}
\usepackage{ulem}
\usepackage{lineno}
\usepackage{amsmath} 
\usepackage{amssymb}
\usepackage{siunitx}
\usepackage{subfig} 
\usepackage{ragged2e}

\newcommand{\cyan}[1]{{\textcolor{cyan}{#1}}}

\newcommand{\LSCO}{La$_{2-x}$Sr$_{x}$CuO$_4$}
\newcommand{\LSCOud}{La$_{1.855}$Sr$_{0.145}$CuO$_4$}

\newcommand{\Qp}{\bf{Q}_{H}^{\pm}}
\newcommand{\Qm}{\bf{Q}_{K}^{\pm}}

\newcommand{\Qone}{\bf{Q}_{2,1}^{\pm}}
\newcommand{\Qthree}{\bf{Q}_{4,3}^{\pm}}

\usepackage{float} 
\usepackage{bibunits}

\begin{document}

 \title{Dynamic Phase Competition: Superconductivity Impeding Spin-Density-Wave Condensation. } 
 \title{Superconductivity Impedes Spin-Density-Wave Condensation: Dynamic Competition. } 
  \title{Spin-Density-Wave Condensation Impeded by Emergence of Superconductivity } 
   \title{Time-Domain Competition: Emergence of Superconductivity Impedes Spin-Density-Wave Condensation} 

 \title{\cyan{Dynamic Competition in a High-Temperature Superconductor}} 
 
\title{\cyan{Dynamic Competition in a magnetically-driven superconductor}} 
\title{\cyan{Symmetry of magnetically-driven Cooper pairs}} 
\title{Dynamic Competition between Cooper-Pair and Spin-Density-Wave Condensation}

   \author{B.~Decrausaz}
 \affiliation{Physik-Institut, Universit\"{a}t Z\"{u}rich, Winterthurerstrasse 190, CH-8057 Z\"{u}rich, Switzerland} 
 \affiliation{PSI Center for Neutron and Muon Sciences, 5232 Villigen PSI, Switzerland} 

 \author{M. Pikulski}
 \affiliation{Laboratory for Solid State Physics, ETH Z\"{u}rich, 8093 Z\"{u}rich, Switzerland}

 \author{O.~Ivashko}
 \affiliation{Physik-Institut, Universit\"{a}t Z\"{u}rich, Winterthurerstrasse 190, CH-8057 Z\"{u}rich, Switzerland} 
 
 \author{N. B.\ Christensen}
 \affiliation{Department of Physics, Technical University of Denmark, DK-2800 Kongens Lyngby, Denmark
 }

 \author{J.~Choi}
 \affiliation{Diamond Light Source, Harwell Campus, Didcot OX11 0DE, United Kingdom}
 \author{L.\ Udby}
 \affiliation{Nano-Science Center, Niels Bohr Institute, University of Copenhagen, DK-2100 Copenhagen, Denmark}
 
 \author{Ch.\  Niedermayer}
  \affiliation{PSI Center for Neutron and Muon Sciences, 5232 Villigen PSI, Switzerland}
 \author{K.\ Lefmann}
 \affiliation{Nano-Science Center, Niels Bohr Institute, University of Copenhagen, DK-2100 Copenhagen, Denmark}
 \author{H. M.\ R\o nnow}
 \affiliation{Laboratory for Quantum Magnetism, \'Ecole Polytechnique F\'ed\'erale de Lausanne (EPFL), CH-1015 Lausanne, Switzerland}
 
 \author{J.\ Mesot}
 \affiliation{PSI Center for Neutron and Muon Sciences, 5232 Villigen PSI, Switzerland}
  \affiliation{Laboratory for Solid State Physics, ETH Z\"{u}rich, 8093 Z\"{u}rich, Switzerland}
 \author{J.\ Ollivier}
 \affiliation{Institut Laue-Langevin, BP 156, F-38042 Grenoble, France}
 \author{T.\ Kurosawa}
 \affiliation{Department of Physics, Hokkaido University - Sapporo 060-0810, Japan}
 \affiliation{Department of Applied Sciences, Muroran Institute of Technology, Muroran 050-8585, Japan}
 \author{N.\ Momono}
 \affiliation{Department of Physics, Hokkaido University - Sapporo 060-0810, Japan}
 \affiliation{Department of Applied Sciences, Muroran Institute of Technology, Muroran 050-8585, Japan}
 \author{M.\ Oda}
 \affiliation{Department of Physics, Hokkaido University - Sapporo 060-0810, Japan}

 \author{J.\ Chang}
\email{johan.chang@physik.uzh.ch}
\affiliation{Physik-Institut, Universit\"{a}t Z\"{u}rich, Winterthurerstrasse 190, CH-8057 Z\"{u}rich, Switzerland} 

 \author{D. G. Mazzone}
\email{daniel.mazzone@psi.ch}
 \affiliation{PSI Center for Neutron and Muon Sciences, 5232 Villigen PSI, Switzerland}
 %
 %
 
 
 \begin{abstract}
Quantum matter phases may co-exist microscopically even when they display competing tendencies. A fundamental question is whether such a competition can be avoided through the elimination of one phase while the other one condenses into the ground state.  Here, we present a high-resolution neutron spectroscopy study of the low-energy spin excitations  in the high-temperature superconductor \LSCOud. In the normal state, we find low-energy magnetic fluctuations at incommensurate reciprocal lattice positions where spin-density-wave order emerges at lower Sr concentration or at high magnetic fields. While these spin excitations are largely suppressed by the emergence of the superconducting spin gap, some low-energy magnetic fluctuations persist deep inside the superconducting state. We interpret this result in terms of a dynamic competition  between superconductivity and magnetism, where superconductivity impedes the condensation of low-energy magnetic fluctuations through the formation of magnetically-mediated Cooper pairs.
 \end{abstract}

 \date{\today}
 
   \maketitle
 \section{Introduction} 

\begin{figure*}
 	\begin{center}
 	\includegraphics[width=0.8\textwidth]{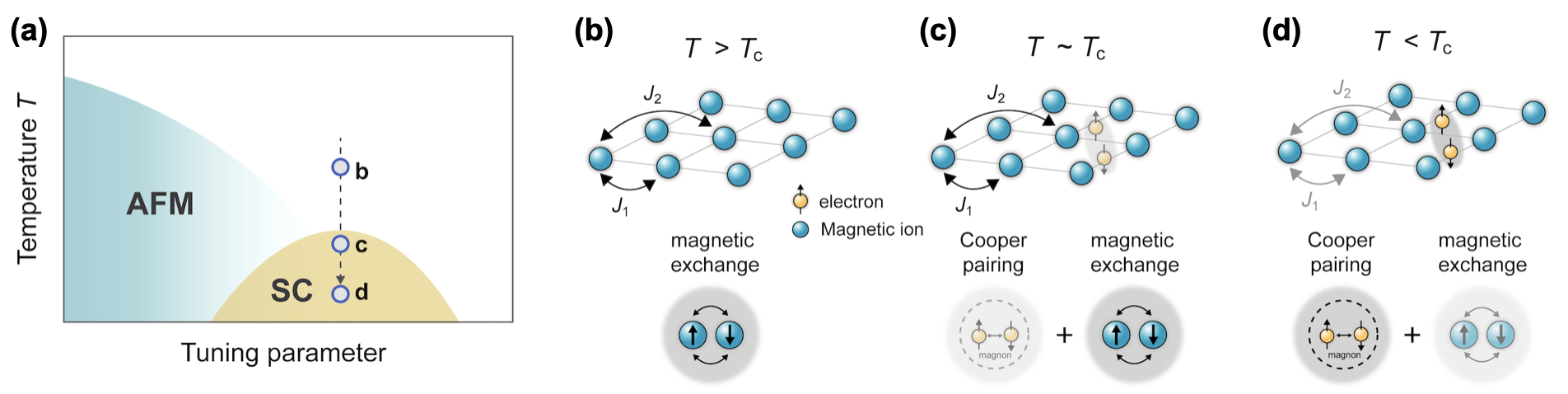}
    \caption{\justifying
        \textbf{(a)} Representative phase diagram of static antiferromagnetism (AFM) competing with unconventional superconductivity. AFM fluctuations persist outside the magnetic long-range ordered phase and are thought to mediate unconventional Cooper pairs~\cite{Monthoux2007}. \textbf{(b)} Above the superconducting transition temperature $T_\text{c}$ the magnetic interactions arise from nearest neighbour exchange couplings $J$ = \{$J_1$, $J_2$,...\}, shaping the magnetic excitation spectrum. Inside the superconducting phase \textbf{(d)} the unconventional Cooper pairs trigger a superconducting spin gap affecting the magnetic excitation spectrum. Here we study the dynamic competition \textbf{(c)} between the magnetic fluctuations and the Cooper pair formation across $T_\text{c}$, by examining the magnetic excitation spectrum via neutron scattering. }
    \label{fig:schematics}
 	\end{center}
 \end{figure*}

After more than a hundred years since the discovery of superconductivity, the field still remains a source of fascination for the scientific community. During the last four decades particular interest in the field arose through the discovery of high-temperature superconductivity.  Many of these unconventional superconducting condensates emerge either in proximity to or in coexistence with magnetically ordered ground states~\cite{MathurNAT1998}. Yet, superconductivity typically competes with static magnetism (see Fig.~\ref{fig:schematics}a)~\cite{NandiPRL2010,ChangPRB2008}. This \textit{static competition} has been explored by a multitude of techniques monitoring the two order parameters or their onset temperatures as a function of temperature or extrinsic tuning parameters including magnetic field, chemical doping and pressure~\cite{Grissonnanche14a,huckerPRB2014,blancocanosaPRB2014}.

While magnetic long-range order tends to compete with superconductivity, it is widely believed that magnetic fluctuations provide the underlying attractive interaction to establish unconventional Cooper pairing~\cite{Scalapino2012}.  This has motivated detailed studies of magnetic excitations in unconventional superconductors using in particular neutron scattering techniques. An experimental signature of superconductivity - found in cuprate, heavy-fermion and iron-based superconductors - is a collective spin-1 excitation below the superconducting transition temperature $T_\text{c}$~\cite{ROSSATMIGNOD1991,TranquadaPRB2004,ChristensenPRL2004,Stock2008,Mazzone2017,Yu2009}. This spectral feature, often referred to as ''resonance``, is thought to arise due to magnetically mediated Cooper pairing as a feedback effect of unconventional superconductivity opening a gap at the Fermi surface. In turn, this is reflected in the spin excitation spectrum by the appearance of a low-energy spin gap, from which the resonance draws its spectral weight \cite{ROSSATMIGNOD1991,Dai2001,Mason1993,YamadaPRL1995}. In the past, several studies have assessed the interplay between the spin resonance and spin gap on the one hand, and static magnetism on the other one. Such studies have yielded insight into the symmetry of the superconducting order parameter and its effect on the magnetic long-range order ~\cite{Mazzone2017,Dai2015}. 
These studies, however, mainly focused on the effects of static order onto the spin resonance, but did not study how the magnetically mediated Cooper pairs affect the dynamic properties which ultimately lead to magnetic long-range order.

 A possible route to address this \textit{dynamic competition} is through studies focusing on how the formation of Cooper pairs affects the low-lying magnetic fluctuations. Notably, unconventional superconductivity often appears close to second-order magnetic phase transitions, where the symmetry breaking magnetic order parameter emerges through a continuous quantum phase transition that is linked to a magnetic soft mode. If unconventional Cooper pairs are built from low-energy magnetic fluctuations, the formation of Cooper pairs is likely impacting the low-energy modes associated to the magnetic long-range order (see Fig.~\ref{fig:schematics}b-d). Thus, insight into this dynamic competition between magnetic fluctuations and unconventional superconductivity can be gained through studies of the low-energy magnetic excitation spectrum. Ideally, such a study is performed either via non-thermal tuning across the static antiferromagnetic (AFM)-superconducting phase boundary~\cite{ChangPRL2009}, or by studying the temperature dependence of the fluctuations in a material without magnetic order close to the AFM phase boundary across $T_\text{c}$ (see Fig.~\ref{fig:schematics}a). 
The latter approach requires a material with a large $T_\text{c}$ to ensure an appreciable temperature window over which the emergent spin gap below the superconducting resonance can be investigated.

  \begin{figure*}
 	\begin{center}
 	\includegraphics[width=0.75\textwidth]{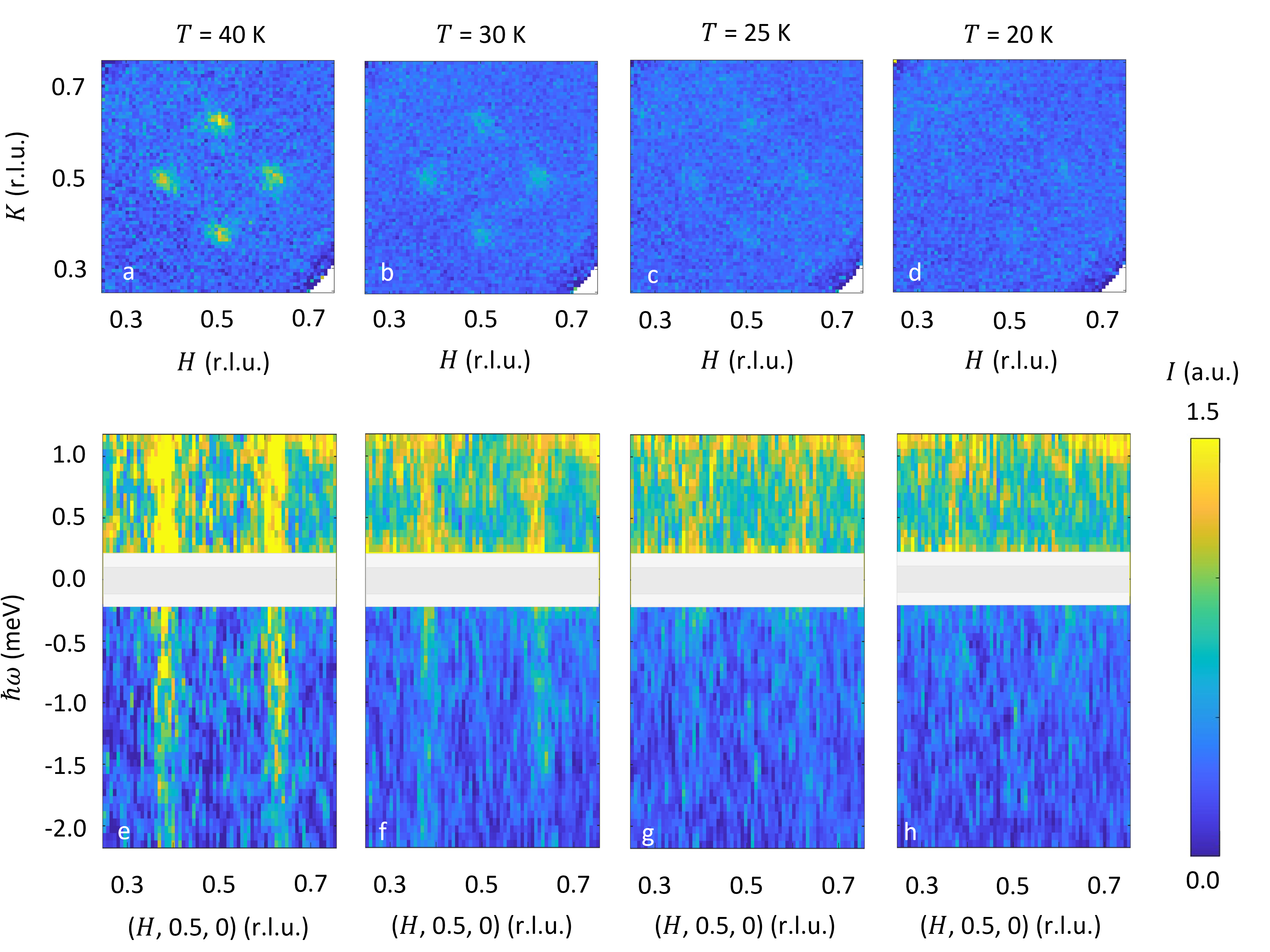}
    \caption{\justifying
    Time-of-flight spectroscopy data using setup II of Table~1. \textbf{(a-d)} Maps in the ($H$,$K$,0)-plane in reciprocal lattice units (r.l.u.) where the inelastic intensity is integrated in the range \mbox{$\hbar\omega$ = [-2.15,-0.15]~meV}.  Four incommensurate spin excitations are clearly visible in the normal state \mbox{($T=40$~K)}. \textbf{(e-h)} Intensity plotted as a function of  \mbox{($H$, 0.5, 0)} and excitation energy with an integration width in $K$ of 0.025~r.l.u.. Negative spin excitation energies indicate that neutrons have gained energy through the scattering process. The elastic signal has been replaced by the shaded area  indicating once and twice the elastic FWHM in dark and light gray, respectively, to improve visibility of the fluctuations in the colorplot.    }
    \label{Qmaps}
 	\end{center}
 \end{figure*}

Here we focus on \LSCOud~which is a model material that fulfills these requirements. It displays no magnetic long-range order at zero magnetic field, and features a superconducting phase below $T_\text{c}$ = 36~K with a spin gap $E_g\simeq $ 4~meV~\cite{ChangPRL2009, ChangPRB2008} appearing at an incommensurate quartet of wavevectors.
Spin stripe-order~\cite{tranquadaEvidenceStripeCorrelations1995} is recovered upon reduction of the Sr content ($x$ $<  $ 0.135)~\cite{ChangPRB2008, KofuPRL2009} and when a magnetic field ($\mu_0$$H$ $>$ 7~T) is a applied along the crystallographic $c$-axis~\cite{ChangPRL2009}. Using time-of-flight neutron spectroscopy with an elastic resolution of $\delta E$ = 80 $\mu$eV full-width at half maximum (FWHM), we observe low-energy magnetic fluctuations for $T \geqq T_\text{c}$ at the wavevectors characteristic of the incipient stripe order. These fluctuations are gradually wiped out by the emergent spin gap, but we observe persistent spectral weight below $\sim$1 meV transfer. The result is interpreted as a near-condensation of low-energy spectral weight that is impeded by the opening of the spin gap. Notably, we observe a dynamic competition where superconductivity prevents spin-density-wave ordering.

 \section{Experimental Details} 

 The experiment was carried out using the IN5 time-of-flight neutron spectrometer~\cite{Ollivier2011} at the 
 Institut Laue-Langevin. A sample of \LSCOud~\cite{crystal}, consisting of two crystals (total mass 3.5~g, $T_\text{c}$ = 36~K), was cut from a single rod grown via the traveling-solvent floating-zone method and co-aligned within less than $1^\circ$. The sample with lattice parameters $a \approx b = 3.81$~\AA, $c \approx 13.2$~\AA~in the high-temperature tetragonal unit cell notation, was measured in a standard cryostat. Magnetic excitation spectra were measured with a fixed sample orientation ($b-$axis perpendicular to the horizontal scattering plane)  for 12 or 24 hours at different temperatures. Two different setups were employed,  with incident wavelengths $\lambda_i = 3.3$ and 5.0~\AA~(see Table~1). The shorter wavelength was used to confirm the size of the superconducting spin gap ($\sim$4~meV)\cite{KofuPRL2009,ChangPRL2009} at the base temperature of $T$ = 2 K (see Supplementary Material (SM) Note 1). The $\lambda$ = 5.0~\AA~ 
setup was used to collect spectra at $T$ = 40, 30, 25 and 20~K
with a resolution of $\delta E$ = 80~$\mu$eV.

  \begin{table}[hbt]
 	\begin{center}
 		\begin{ruledtabular}
 			\begin{tabular}{cccc}
 				Setup &	$\lambda_i$ (\AA)  & $\delta E$ ($\mu$eV) & $\omega$ interval (meV)  \\\hline
 				I &	3.3 & 384 & $-3\rightarrow-0.4$ and $0.4\rightarrow6$ \\
 				II&	5.0 &  80 & $-3\rightarrow-0.1$ and $0.1\rightarrow1.5$ \\
 				
 			\end{tabular}
 		\end{ruledtabular}
 		\caption{Incident neutron wavelengths $\lambda_i$ and corresponding 
 			elastic resolution $\delta E$ (FWHM). The high-resolution setting (setup II) enabled a study of the low-energy spin-excitations on both the neutron gain and loss side. With setup I, we confirmed the size of the superconducting spin gap\cite{KofuPRL2009,ChangPRL2009}  at base temperature (see SM Note 1).}
 	\end{center}
    \label{tab:tab1}
 \end{table} 

   \begin{figure*}
 	\begin{center}
 		\includegraphics[width=\textwidth]{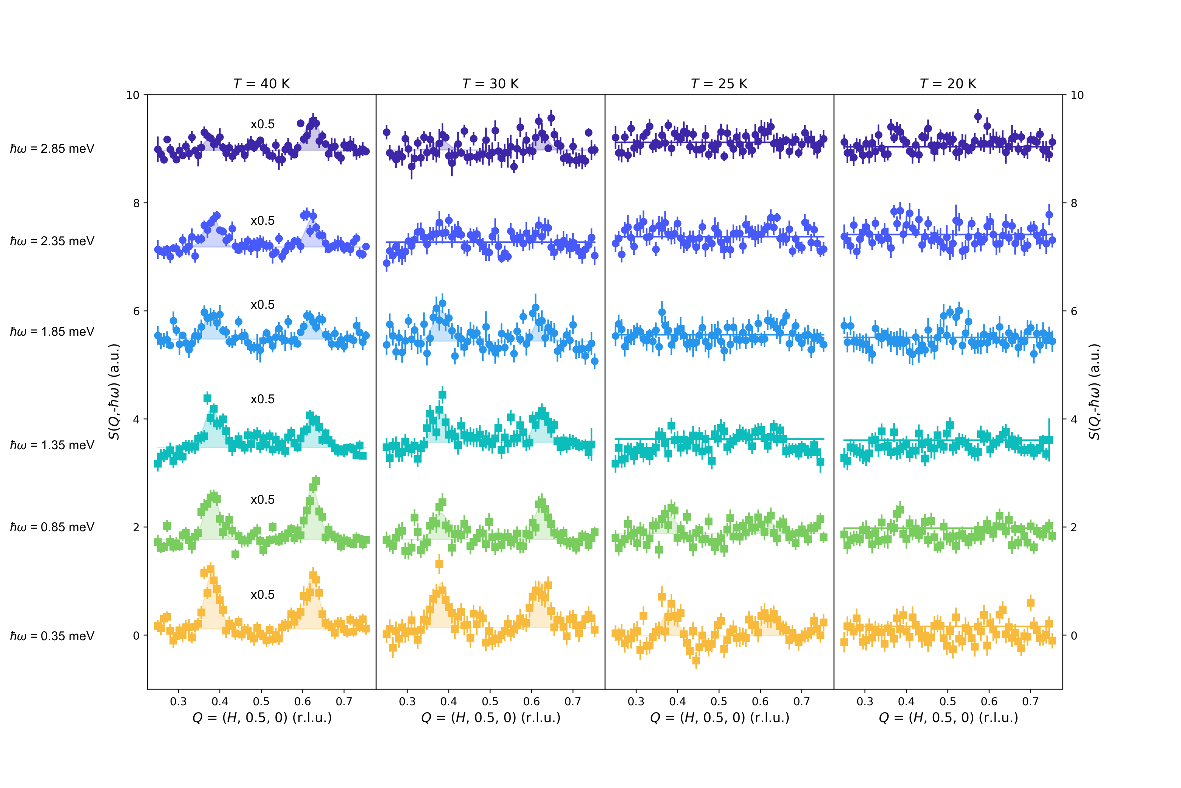}
 		\caption{\justifying
            Dynamic structure factor $S(\bf{Q},-\hbar\omega)$ at different energy transfers and temperatures. Here $\bf{Q}$ \mbox{= ($H$, 0.5, 0)} represents a reciprocal space cut through the spin excitations for which the intensity of the $H$ and $K$ cuts were combined. The chosen integration widths were $\pm$0.1~meV along the energy axis and 0.5 $\pm$0.125 (r.l.u.) in the perpendicular $\bf{Q}$-direction. For presentation purposes each of the $Q$-scans was shifted along the $y$-axis. For the energies \mbox{$\hbar\omega$ = 1.35}, 0.85 and 0.35~meV detailed balance was used to take advantage of both energy gain and loss data. Further details about the analysis can be found in the SM.}
 		\label{fig:fig2}
 		
 	\end{center}
 \end{figure*}

\section{Results} 
  
The neutron scattering rate is proportional to the four-dimensional dynamic structure factor $S(\bf{Q},\hbar\omega)$, which is directly connected to the dynamic magnetic susceptibility: $S(\bf{Q},\hbar\omega)\propto\chi^{''}(\bf{Q},\hbar\omega)$ with a proportionality factor equal to the Bose occupation factor $n_\text{B}(\hbar\omega,T)$ = 1/[exp($\hbar\omega/k_\text{B}T$)-1] for neutron energy-gain processes ($\hbar\omega<0$) and to $(n_\text{B}(\hbar\omega,T)+1)$ for energy-loss processes ($\hbar\omega>0$). Here, $\bf{Q}$ represents the three dimensional momentum transfer, $\hbar\omega$ the energy transfer and $k_\text{B}$ the Boltzmann constant. As the electronic band structure and spin excitations of \LSCO~are quasi two-dimensional~\cite{HorioPRL2018,RoemerPRB2015,Lake2005}, momentum transfers $\bf{Q}=$($H$,$K$,$L$) were projected to ($H$,$K$,0)\cite{ChristensenPRL2004,VignolleNATPHYS2007,MonneyPRB2016}. Figure~\ref{Qmaps}a-d illustrates the resulting scattered neutron intensity of \LSCOud~in the $HK$-plane, integrated over an energy range of \mbox{$\hbar\omega$ = [-2.15,-0.15]~meV} for $T$ = 40, 30, 25 and 20~K. A step size of $\delta Q = $ 0.0075~r.l.u. was chosen for this plot. The $\bf{Q}$-map at $T$ = 40~K reveals an enhanced neutron intensity at wavevectors  $\Qp$=($1/2 \pm
 \delta$,1/2,0)  and $\Qm$=($1/2$,$1/2 \pm \delta$,0) with $\delta\approx0.12$. These excitations, known to arise from the magnetic interactions in \LSCO~\cite{Dean2013}, appear as straight columns in the measured energy range because the spin-wave bandwidth amounts to$~\sim$300~meV. Their reciprocal lattice positions coincide with the positions where magnetic Bragg peaks appear at slightly lower Sr doping or under magnetic field~\cite{ChangPRB2008,ChangPRL2009}. Thus, we interpret the excitations as signatures of the nearby ordered state.

When the temperature is reduced below the superconducting transition temperature ($T_\text{c}$ = 36~K) the magnetic excitation spectrum is gradually modified (see Fig.~\ref{Qmaps}). 
 Detailed insight into the temperature dependence of the low-energy fluctuations is gained in Fig. \ref{Qmaps}e-h, where the energy dependence of the magnetic fluctuations is shown along ($H$,0.5,0). Here, we used an integration width along (0, 0.5 $\pm$0.025, 0) ~r.l.u, a step sizes $\delta Q$ along $H$ and 60~$\mu$eV along the energy axis. The shaded area extends over one and two elastic FWHMs, colored in dark and light gray, respectively. The data at $T$ = 40~K shows two excitation branches at $\Qp$=($1/2 \pm\delta$,1/2,0) with energy dependent intensity modulation peaking below an energy transfer of $\sim$1~meV. Upon cooling below $T_\text{c}$ the magnetic intensity is gradually suppressed, but our results suggest that magnetic fluctuations persist at low energy transfers and remain visible down to $T$ $\approx$ 20~K (c.f. Fig.~\ref{Qmaps}g).
 
An analysis of this behavior was carried out through $\bf{Q}$-cuts at constant energy transfers, using an optimized integration width (0.025~r.l.u.) and step size $\delta Q$ to obtain the best signal-to-background ratio (see SM Note 2). Data at various energy transfers $\hbar\omega$ were integrated over $\pm$0.1~meV intervals. Cuts along  ($H$,1/2,0)  and (1/2,$K$,0) for different $\hbar\omega$ and temperatures were fitted with two Gaussians on a constant background. Since no significant variances between cuts along $H$ and $K$ were observed (see SM Note 3), we  combined $H$ and $K$ scans to produce Fig.~\ref{fig:fig2}.  We furthermore used detailed balance, $S(-\bf{Q},-\hbar\omega)$= exp($-\hbar\omega/k_\text{B}T$) $S(\bf{Q},\hbar\omega$), to combine energy-loss and energy-gain spectra (see SM Note 4). Figure~\ref{fig:fig2} displays the dynamic structure factor $S(\bf{Q},-\hbar\omega)$ for different temperatures across the excitation branches at various excitation energies. We note that the $\bf{Q}$-cuts have been offset for clarity. At $T$ = 40~K, two Gaussian peaks at $\Qp$=($1/2 \pm\delta$,1/2,0) are observed over the entire probed energy range. The magnetic fluctuations are significantly weakened below $T_\text{c}$ = 36~K but remain visible up to $\sim$2~meV at $T$ = 30~K. Deeper inside the superconducting phase at $T$ = 25~K, the two excitation branches are observed below $\sim$1~meV. These two excitations are largely suppressed above 0.35~meV at $T$ = 20~K. 

    \begin{figure}
   	\begin{center}
   		\includegraphics[width=0.49\textwidth]{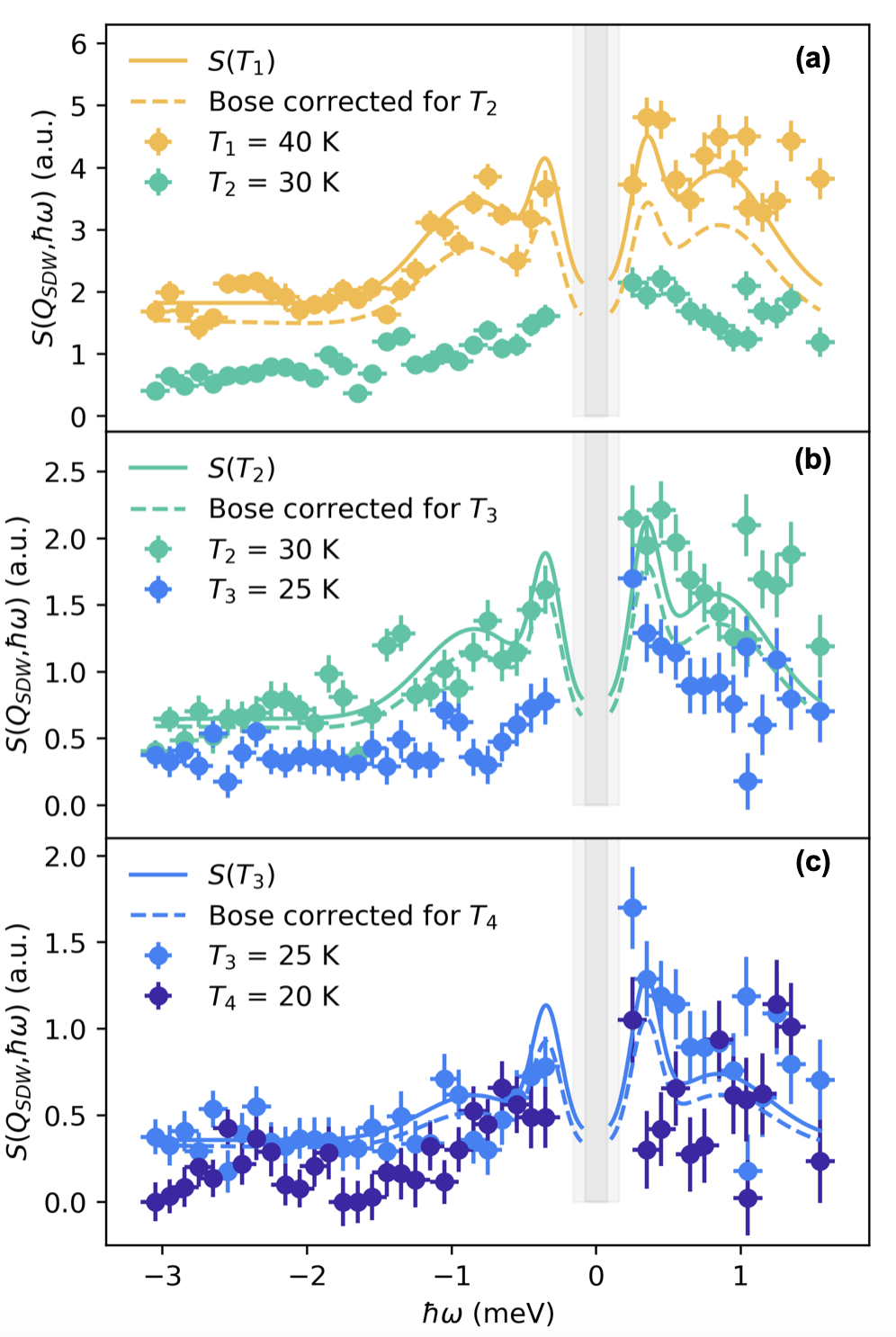}
   		 \caption{\justifying
            Dynamic structure factor as a function of excitation energy for different temperatures. Negative (positive) values indicate that neutrons having gained (lost) energy in the scattering process. The dark (light) gray area symbolises once (twice) the resolution of the instrument. \textbf{(a-c)} compare pair-wise structure factors for temperatures 40, 30, 25 and 20~K. The solid lines are guides to the eye, consisting of four Gaussian lineshapes with a constant background. The two Gaussians on the energy loss side are constrained to the ones on the energy gain side using detailed balance $S(-\bf{Q},-\hbar\omega)$= exp($-\hbar\omega/k_\text{B}T$)$S(\bf{Q},\hbar\omega)$. The dashed lines simulate the expected signal reduction of the solid lines if the temperature dependence is attributed to the Bose occupation factor only.}
\label{fig:fig3}
  	\end{center}
   \end{figure}

This non-uniform energy dependence of the magnetic fluctuations is shown in Fig.~\ref{fig:fig3} for the different temperatures. Here, the data  at $\Qp$=($1/2 \pm\delta$,1/2,0) and $\Qm$=($1/2$,$1/2 \pm \delta$,0) were combined. At each excitation energy and temperature, the  $\bf{Q}$-cuts were fitted to Gaussian lineshapes, and the average amplitude at the peak center $\bf{Q}_{SDW}$ is represented as one point in Fig.~\ref{fig:fig3} (see SM Note 5 for details). The subpanels compare pair-wise $S(\bf{Q}_{SDW},\hbar\omega)$ for $T$ = 40, 30, 25 and 20~K. They confirm the behavior observed in Fig.~\ref{fig:fig2} that low-energy magnetic fluctuations persist to low temperature and that the overall magnetic spectral weight is suppressed (see also SM Note 6). As guides to the eyes the higher temperature $S(\bf{Q}_{SDW},\hbar\omega)$ of each subpanel was fitted to an effective spectral lineshape consisting of four Gaussians on a constant background (see solid lines in Fig.~\ref{fig:fig3}), where the two Gaussians on the neutron energy-loss side were constrained to the ones on the energy-gain side using detailed balance. The dashed lines in the panels simulate the expected variation of the solid line in a scenario when the temperature dependence is attributed to  the Bose occupation factor only. This unambiguously demonstrates that the observed suppression of the low-energy magnetic fluctuations at $T$ = 30, 25 and 20~K does not arise alone from thermal depopulation, but is connected to the another effect impacting the dynamic susceptibility.

\section{Discussion}

 The opening of a superconducting spin gap provides a natural explanation for the suppression of low-energy fluctuations \cite{ChangPRL2009,KofuPRL2009,LakeNAT2002}. However, by analogy to a gap in the electronic spectrum one would expect that, as the sample is cooled below $T_c$ the spin gap should initially be detected at the lowest energy transfers, and that higher-energy excitations should only be wiped out at lower temperatures. Instead, we observe that the magnetic fluctuations are suppressed equally over an energy window of $\sim$3~meV (see SM Note 6 for details), and that low energy fluctuations below $\sim$1~meV persist over an appreciable temperature window down to roughly 20~K. Thus, our results indicate the presence of enhanced low-energy spectral weight in \LSCOud, as indicated by the Gaussian fits in Fig.~\ref{fig:fig3}.  
 
In a study~\cite{KofuPRL2009} of the low-energy fluctuations in \LSCO~over the composition-range $x=0.12-0.135$ a two-component spectral response bearing similarity to our observations was observed at base temperature in samples displaying magnetic order. Here, it was hypothezised that stripe-ordered domains with associated low-energy glassy fluctuations coexist with superconducting regions, displaying a spin gap of order $E_g=1.5~k_BT_c$ \cite{KofuPRL2009}. Following this interpretation, the low-energy components of the temperature-dependent spectrum observed in \LSCOud\ correspond to critical fluctuations of the incipient stripe-ordered phase. Because \LSCOud\ resides on the disordered side of the stripe quantum critical point these slow fluctuations eventually succumb to the opening of the spin gap, while they survive in the magnetically ordered samples studied in Ref. \onlinecite{KofuPRL2009}. 

The interpretation of the persistent low-energy fluctuations as the harbingers of a proximate ordered state, at first sight appears at odds with the prior observation by some of us that the application of a magnetic field perpendicular to the CuO$_2$ planes results in a gradual decrease of the spin gap, and that magnetic order appears when this gap closes \cite{ChangPRL2009}. We remark, however, that vortices may not only suppress superconductivity and with it the spin gap, but could also facilitate a slowing down of the low-energy critical fluctations. This could have been missed in Ref. \onlinecite{ChangPRL2009} because the energy resolution was significantly worse than what we obtained in the present experiment. Thus, further high-resolution experiments on other \LSCO~compositions will be fruitful to check this hypothesis, and to clarify the nature of the low-energy fluctuations below $\sim$1~meV.

In a more general context we notice that cases with two coexistent static order parameters displaying competing tendencies are usually referred to as phase competition. Theoretically such systems are often described by a single complex-valued order parameter~\cite{BrussPRB1999,Zhang1997},
where the competing interaction provides an adiabatic continuity between the two phases (see Fig.~\ref{fig:schematics}a). Our study suggests that the phase competition can be extended beyond the static co-existence phase. When only one phase has condensed into the ground state (in this case superconductivity), a dynamic phase competition indicates a state where the fluctuations of the other order parameter shows competing tendencies. \LSCOud~features at least three non-thermal routes through extrinsic tuning to static AFM order~\cite{KhaykovichPRB2005,ChangPRB2008,ChangPRL2009,LakeNAT2002}.  Thus, the AFM paramagnetic fluctuations are about to condense at $T=40$~K, just above the superconducting transition. As the superconducting gap opens, particle-hole excitations are suppressed, leading to a competition between the condensation of unconventional Cooper-pairs and AFM fluctuations. In \LSCOud~superconductivity dominates this competition, but extrinsic tuning shows an adiabatic connection to static magnetic order where the magnetic modes soften into the ground state. Notably, the low energy scale of the observed enhanced spectral weight below 1~meV may explain why only small perturbations (Sr concentration $<$ 0.135 and $\mu_0H >$ 7~T) induces magnetic long-range order.

The concept of a dynamic competition between superconductivity and magnetism is likely existing also in other materials. In the heavy-fermion materials such as CeCu$_2$Si$_2$ for instance, magnetism and superconductivity possess very similar onset temperatures~\cite{StockertNATPHYS2011}. The spin gap ($\sim$0.3 meV) is, however, an order of magnitude smaller than in \LSCOud~\cite{StockertNATPHYS2011}, which in turn suggests a dynamic competition on a much smaller energy scale. For many pnictides, AFM order sets in above superconductivity, leading to a static competition between magnetism and superconductivity~\cite{NandiPRL2010}. However, close to the magnetic quantum critical point, dynamic competition should arise. For instance, in Ba$_{0.6}$K$_{0.4}$Fe$_2$As$_2$, the spin resonance is sufficiently high in excitation energy enabling future investigations of dynamic competition between magnetism and superconductivity with neutron spectroscopy~\cite{ChristiansonNAT2008}.

\section{Outlook}
Finally, it is interesting to consider dynamic competitions between quantum phases beyond the interplay between magnetism and superconductivity. The field of quantum materials features a plethora of systems with competing phases, but in many cases a dynamic competition is avoided. In some cases, associated fluctuations manifest at different wavevectors. This is the case in CeCu$_{5.8}$Ag$_{0.2}$, for instance~\cite{Poudel2019}. In this material spin-density wave-like fluctuations condensate into the ground state establishing magnetic order at higher Ag concentration, while the competing fluctuations remain gapped. The superconducting ground state is special in this regard, because the Cooper pair condensation leads to a gap over an appreciable range in reciprocal space which easily impacts other fluctuations. Thus, dynamic competition may be expected also in the interplay between superconductivity and charge order, and materials featuring spin- and charge-density fluctuations. Future experiments could be beneficial to better understand the concept of dynamic competition between different quantum phases, and compare it to cases - such as frustrated magnetic materials - where a a dynamic cooperation among low-energy fluctuations is established \cite{Balents2010,Broholm2020}.\\[3mm]

  \textbf{Acknowlegdements:} The time-of-flight neutron experiments were performed at the spectrometer IN5~\cite{Ollivier2011} of the Institut Laue-Langevin,
 France. The experimental data used in this work can be found using the link \href{https://doi.org/10.5281/zenodo.10992720}{https://doi.org/10.5281/zenodo.10992720}. We acknowledge beam line support and technical advices from H. Mutka.  This work was supported by the Swiss National Science Fundation (through NCCR, MaNEP, and grant Nr PZ00P2-142434) and the Ministry
 of Education and Science of Japan.

\bibliography{refv5,Zotero-References}


\clearpage
\begin{bibunit}[apsrev4-1]
\makeatletter
\def\@sect#1#2#3#4#5#6[#7]#8{%
  \ifnum #2>\c@secnumdepth
    \let\@svsec\@empty
  \else
    \refstepcounter{#1}%
    \protected@edef\@svsec{\@seccntformat{#1}\relax}%
  \fi
  \@tempskipa #4\relax
  \ifdim \@tempskipa>\z@
    \begingroup #6\relax
      \@hangfrom{\hskip #3\relax\@svsec}{#8\par}%
    \endgroup
  \else
    \def\@svsechd{#6\hskip #3\relax\@svsec #8\csname #1mark\endcsname{#7}%
                  \addcontentsline{toc}{#1}{%
                    \ifnum #2>\c@secnumdepth \else
                      \protect\numberline{\csname the#1\endcsname}%
                    \fi
                    #7}}%
  \fi
  \@xsect{#5}}

\makeatother

\onecolumngrid
\setcounter{page}{1}        

\newcommand{\beginsupplement}{
        \setcounter{table}{0}
        \renewcommand{\thetable}{S\arabic{table}}
        \setcounter{figure}{0}
        \renewcommand{\figurename}{\textbf{Supplementary Figure}}
}


\begin{center}
    {\large \bfseries Supplementary Material for: Dynamic Competition between Cooper-Pair and Spin-Density-Wave Condensation \par}
    \vspace{1em}

    {\normalsize
    B.~Decrausaz$^{1,2}$, M.~Pikulski$^{3}$, O.~Ivashko$^{1}$, N.B.~Christensen$^{4}$, J.~Choi$^{5}$, L.~Udby$^{6}$,\\
    Ch.~Niedermayer$^{2}$, K.~Lefmann$^{6}$, H.M.~R\o nnow$^{7}$, J.~Mesot$^{2,3}$, J.~Ollivier$^{8}$, T.~Kurosawa$^{9,10}$,\\
    N.~Momono$^{9,10}$, M.~Oda$^{9}$, J.~Chang$^{1}$\footnote{Email: johan.chang@physik.uzh.ch}, D.G.~Mazzone$^{2}$\footnote{Email: daniel.mazzone@psi.ch}
    \par}
    \vspace{0.5em}

    {\footnotesize
    \textit{$^1$Physik-Institut, Universität Zürich, Winterthurerstrasse 190, CH-8057 Zürich, Switzerland}\\
    \textit{$^2$PSI Center for Neutron and Muon Sciences, 5232 Villigen PSI, Switzerland}\\
    \textit{$^3$Laboratory for Solid State Physics, ETH Zürich, 8093 Zürich, Switzerland}\\
    \textit{$^4$Department of Physics, Technical University of Denmark, DK-2800 Kongens Lyngby, Denmark}\\
    \textit{$^5$Diamond Light Source, Harwell Campus, Didcot OX11 0DE, United Kingdom}\\
    \textit{$^6$Nano-Science Center, Niels Bohr Institute, University of Copenhagen, DK-2100 Copenhagen, Denmark}\\
    \textit{$^7$Laboratory for Quantum Magnetism, EPFL, CH-1015 Lausanne, Switzerland}\\
    \textit{$^8$Institut Laue-Langevin, BP 156, F-38042 Grenoble, France}\\
    \textit{$^9$Department of Physics, Hokkaido University, Sapporo 060-0810, Japan}\\
    \textit{$^{10}$Department of Applied Sciences, Muroran Institute of Technology, Muroran 050-8585, Japan}
    \par}

    \vspace{0.5em}
    {\footnotesize (Dated: \today)}
\end{center}

\vspace{1em}

\section*{\texorpdfstring{S\MakeLowercase{upplementary} N\MakeLowercase{ote}}{Supplementary Note} 1. C\MakeLowercase{onfirmation of the }S\MakeLowercase{uperconducting} S\MakeLowercase{pin} G\MakeLowercase{ap}}

\setcounter{figure}{0}

\renewcommand{\figurename}{\textbf{Supplementary Figure}}

\renewcommand{\thefigure}{\arabic{figure}}

\begin{figure}[H]
 	\begin{center}
 	\includegraphics[width=1\textwidth]{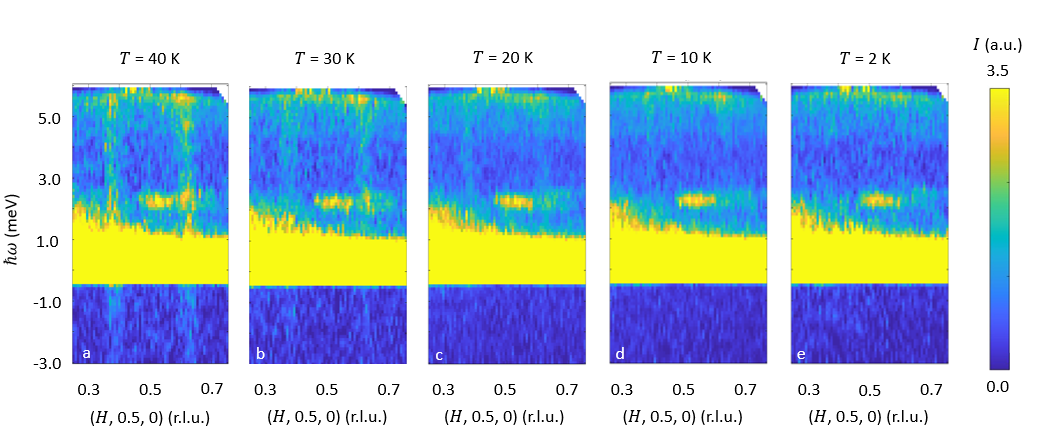}
        \caption{Time-of-flight spectroscopy data using $\lambda_i = 3.3$~\AA~. The Intensity is plotted as a function of  \mbox{($H$, 0.5, 0)} in r.l.u and excitation energy with an integration width in $K$ of 0.025~r.l.u. for temperatures \textbf{(a)} $T$ = 40~K $>$ $T_\text{c}$, \textbf{(b)} $T$ = 30~K, \textbf{(c)} $T$ = 20~K, \textbf{(d)} $T$ = 10~K,and \textbf{(e)} $T$ = 2~K . Negative spin excitation energies indicate that neutrons have gained energy through the scattering process.}
        \label{Fig:whole gap 1}
 	\end{center}
 \end{figure}

\begin{figure}[H]
 	\begin{center}
 	\includegraphics[width=0.8\textwidth]{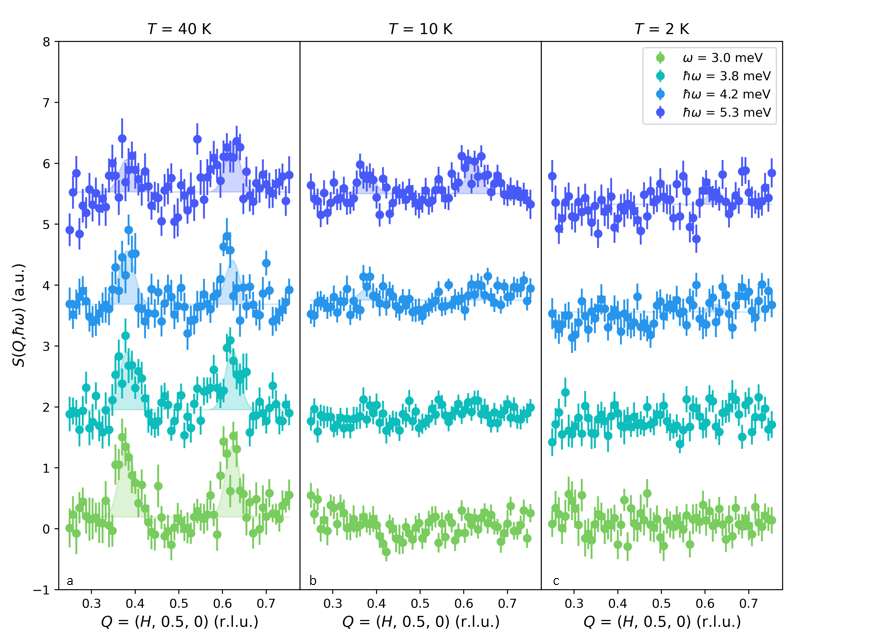}
        \caption{Time-of-flight spectroscopy data using $\lambda_i = 3.3$~\AA~. The dynamic structure factor $S(\bf{Q},\hbar\omega)$ at different energy transfers is plotted for the temperatures $T$ = 40~K $>$ $T_\text{c}$, $T$ = 10~K and $T$ = 2~K . Here $\bf{Q}$ \mbox{= ($H$, 0.5, 0)} represents a reciprocal space cut through the spin excitations for which the intensity of the $H$ and $K$ cuts were combined. An integration width of $\pm$0.1~meV was used along the energy axis, and over 0.5 $\pm$0.125 (r.l.u.) in the perpendicular $\bf{Q}$-direction. For presentation purposes each of the $Q$-scans was shifted along the $y$-axis. Solid lines are Gaussian fits.}
        \label{Fig:whole gap 2}
 	\end{center}
 \end{figure}

A setup with an incoming neutron wavelength of $\lambda_i = 3.3$~\AA{} was used to confirm the size of the superconducting spin gap. The resulting intensity is depicted in Fig.~\ref{Fig:whole gap 1} as function of \mbox{($H$, 0.5, 0)} in r.l.u and excitation energy for various temperatures Fig.~\ref{Fig:whole gap 2} presents one-dimensional cuts through the data along \mbox{= ($H$, 0.5, 0)} for different energies. Our observations
are consistent with $E_g$ $\approx$ 4~meV found previously\cite{KofuPRL2009,ChangPRL2009}.
In Fig.~\ref{Fig:whole gap 1}a and Fig.~\ref{Fig:whole gap 2}a, we show the spin excitations in the normal state, at $T$ = 40~K $>$ $T_\text{c}$. Incommensurate excitations are observed down to the lowest measurable excitation energies. At base temperature $T$ = 2~K the intensity was to low to extract the gap size because of the Bose factor reduction (see Fig.~\ref{Fig:whole gap 2}e). However, at $T$ = 10~K the spectral weight up to $\hbar\omega$ $\approx$ 4~meV is gapped out. The enhanced intensity near $\hbar\omega$ $\sim$2 meV in Fig.~\ref{Fig:whole gap 1} is attributed to parasitic scattering from the IN5 sample environment.

\section*{\texorpdfstring{S\MakeLowercase{upplementary} N\MakeLowercase{ote}}{Supplementary Note} 2. O\MakeLowercase{ptimized} I\MakeLowercase{ntegration} W\MakeLowercase{idth in} $Q$ }
The four dimensional $S(\bf{Q},\hbar\omega)$ data obtained from IN5 was reduced to one dimension, integrating over the other three dimensions. Along $L$ the data was integrated over the full range covered by the data set  as the spin excitations in LSCO are assumed to be two dimensional  \cite{tranquadaNeutronscatteringStudyStripephase1996,lakeThreedimensionalityFieldinducedMagnetism2005}. To assess the optimal integration width along $H$ and $K$ we first investigated various fixed energy integration widths at different temperatures (see Table \ref{table T vs E}). They were chosen to assess whether a single $\bf{Q}-$integration width can be used over the entire energy range. We note that the 20~K data were neglected during this procedure due to the weak signal.

\begin{figure}[H]
    \centering
    \subfloat[]{%
        \includegraphics[width=0.45\textwidth]{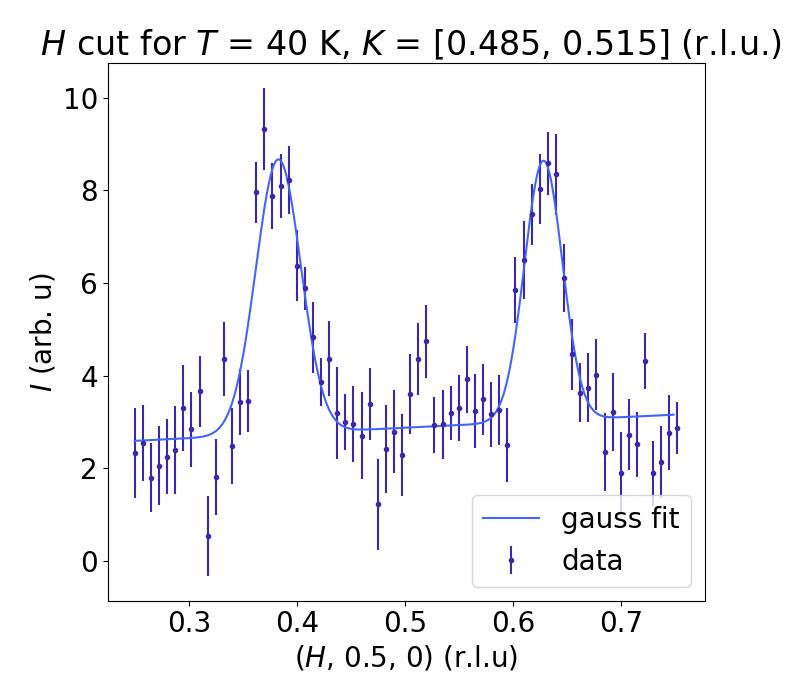}
        \label{h cut}
    }
    \hfill
    \subfloat[]{\includegraphics[width=0.45\textwidth]{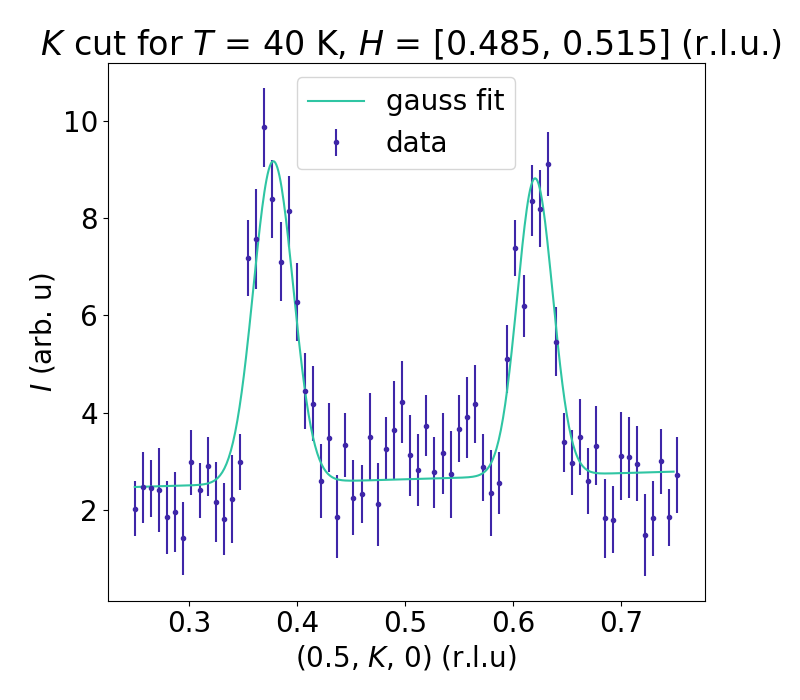}
    \label{k cut}}
    \caption{Examples of $H$ and $K$ cuts. Constant energy cut
 at 40~K with $E$ = [-1.9, 0.9]~(meV) and an integration width along $Q$ of 0.03~(r.l.u.)  in \textbf{(a)} along ($H$ 0.5 0) and in \textbf{(b)} along (0.5 $K$ 0). The fitted Gaussian lineshape is indicated by the solid line. }
    \label{fig:HK_cuts}
\end{figure}

\begin{figure}[H]
    \centering
    \subfloat[]{\includegraphics[width=0.45\textwidth]{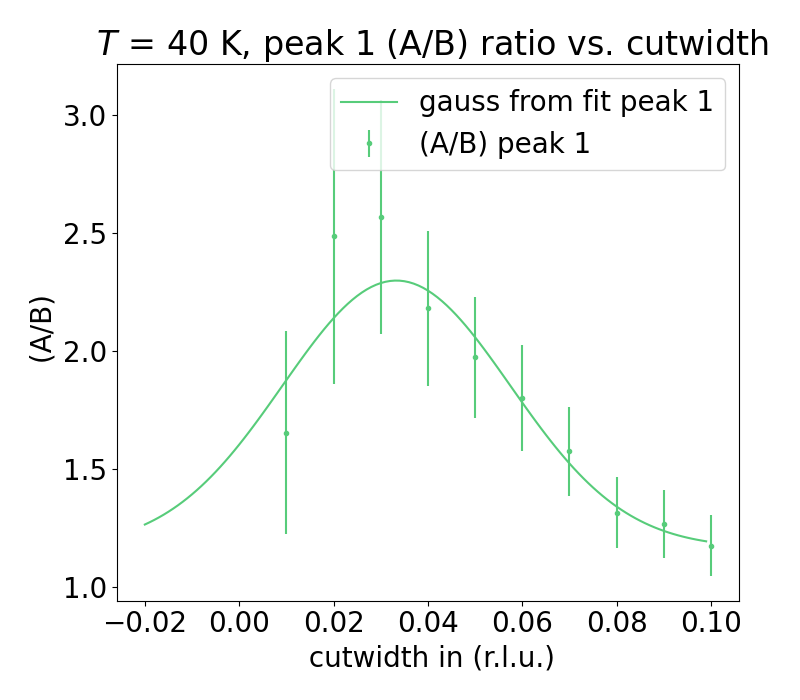}
    \label{A/B 40K peak 1}}
    \hfill
    \subfloat[]{\includegraphics[width=0.45\textwidth]{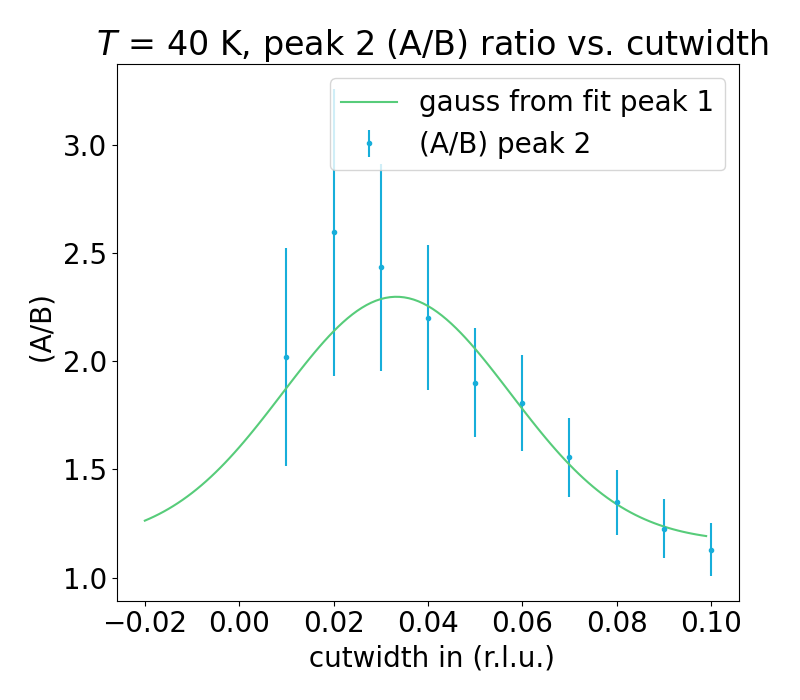}
    \label{A/B 40K peak 2}}

    \subfloat[]{\includegraphics[width=0.45\textwidth]{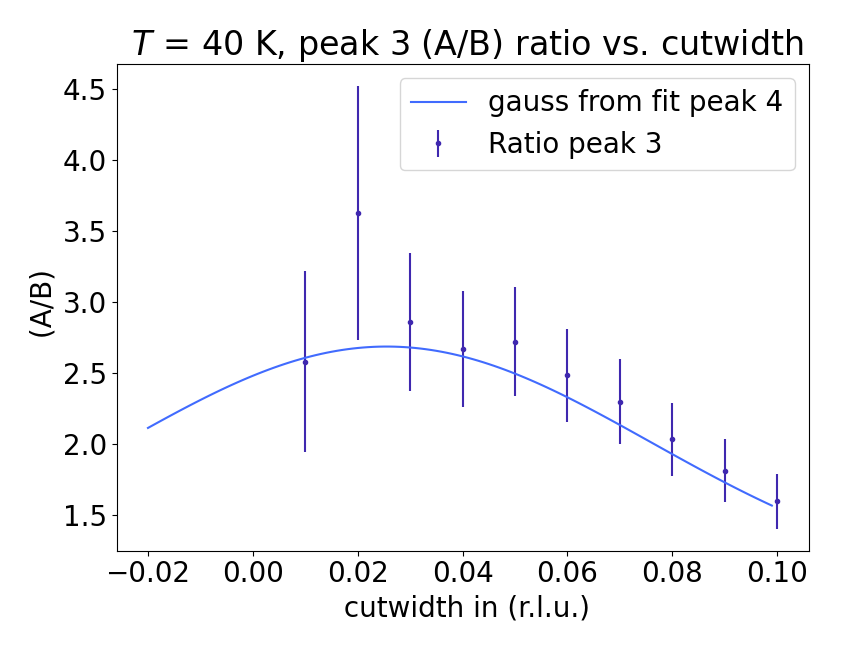}
    \label{A/B 40K peak 3}}
    \hfill
    \subfloat[]{\includegraphics[width=0.45\textwidth]{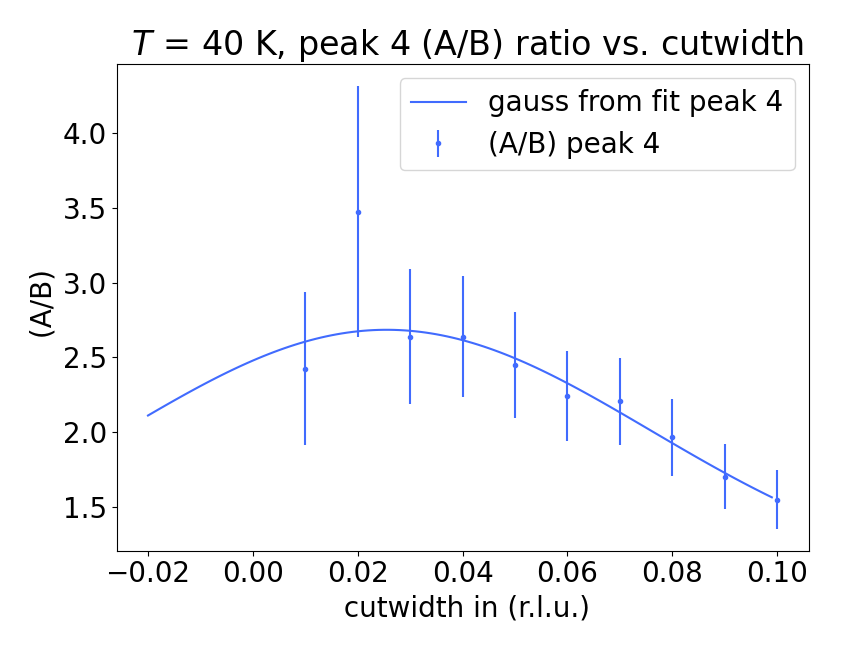}
    \label{A/B 40K peak 4}}
    
    \caption{Determination of the optimal integration width in $q$ for the 40~K data. The amplitude-to-background ratio (A/B) and its dependence on the momentum integration width were examined for peaks 1-4 located at $\Qone$=($1/2 \pm \delta$,1/2,0)  and $\Qthree$=($1/2$,$1/2 \pm \delta$,0) with $\delta\approx0.12$. The optimal integration width was determined to the value where A/B no longer improves. The result was also fitted with a Gaussian lineshape for peaks 1 and 4. The fit was overlayed with the data of peaks 2 and 3 exhibiting  the same optimal integration width in $q$.}
    \label{A/B 40K}
\end{figure}

\begin{figure}[H]
    \centering
    \subfloat[]{\includegraphics[width=0.45\textwidth]{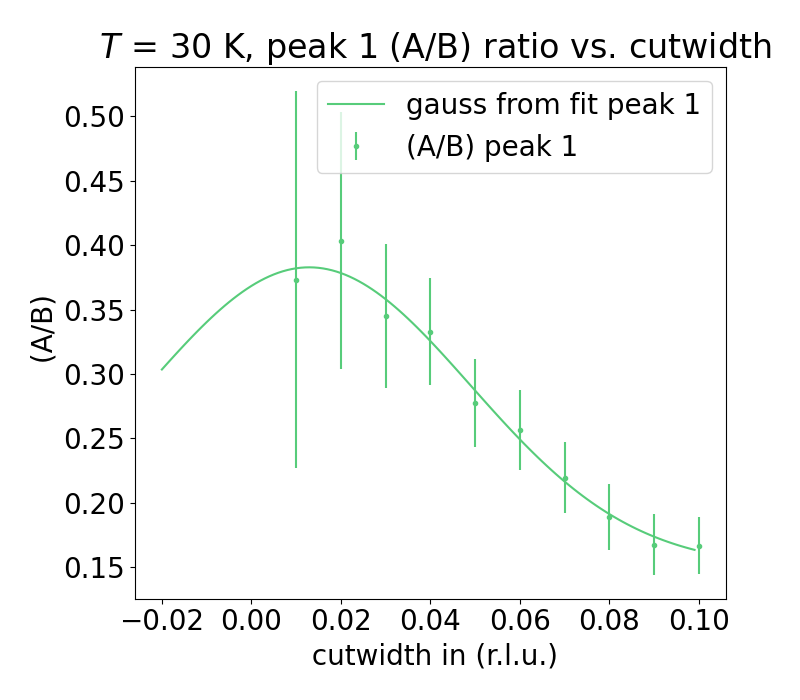}
    \label{A/B 30K peak 1}}
    \hfill
    \subfloat[]{\includegraphics[width=0.45\textwidth]{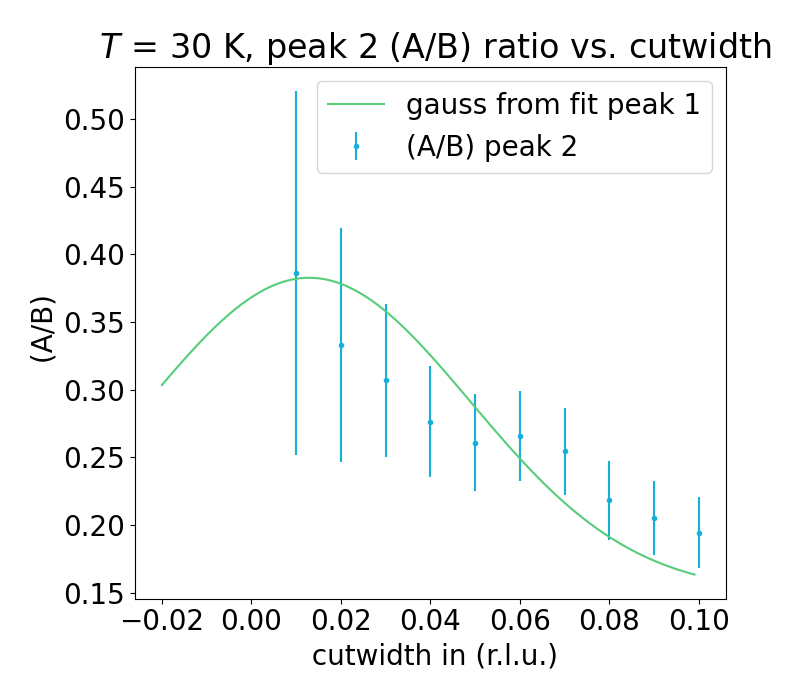}
    \label{A/B 30K peak 2}}

    \subfloat[]{\includegraphics[width=0.45\textwidth]{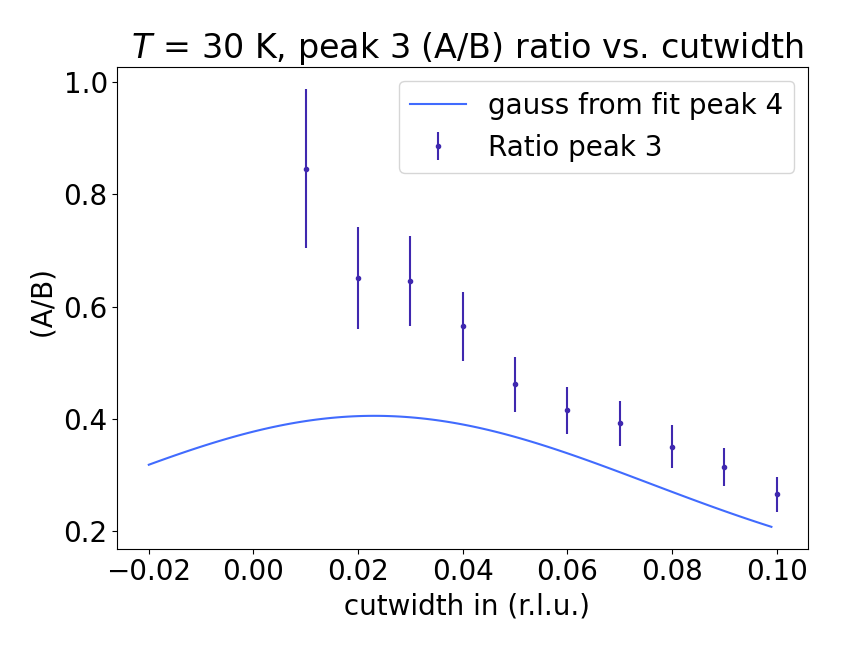}
    \label{A/B 30K peak 3}}
    \hfill
    \subfloat[]{\includegraphics[width=0.45\textwidth]{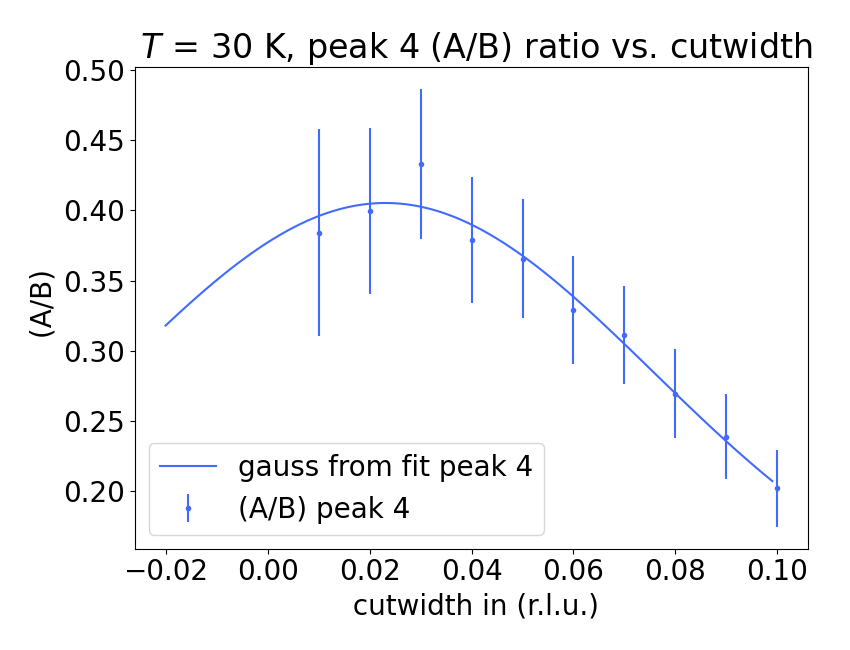}
    \label{A/B 30K peak 4}}
    
    \caption{Determination of the optimal integration width in $q$ for the 30~K data. The amplitude-to-background ratio (A/B) and its dependence on the momentum integration width were examined for peaks 1-4 located at $\Qone$=($1/2 \pm \delta$,1/2,0)  and $\Qthree$=($1/2$,$1/2 \pm \delta$,0) with $\delta\approx0.12$. The optimal integration width was determined to the value where A/B no longer improves. The result was also fitted with a Gaussian lineshape for peaks 1 and 4. The fit was overlayed with the data of peaks 2 and 3 exhibiting  the same optimal integration width in $q$.}
    \label{A/B 30K}
\end{figure}

\begin{figure}[H]
    \centering
    \subfloat[]{\includegraphics[width=0.45\textwidth]{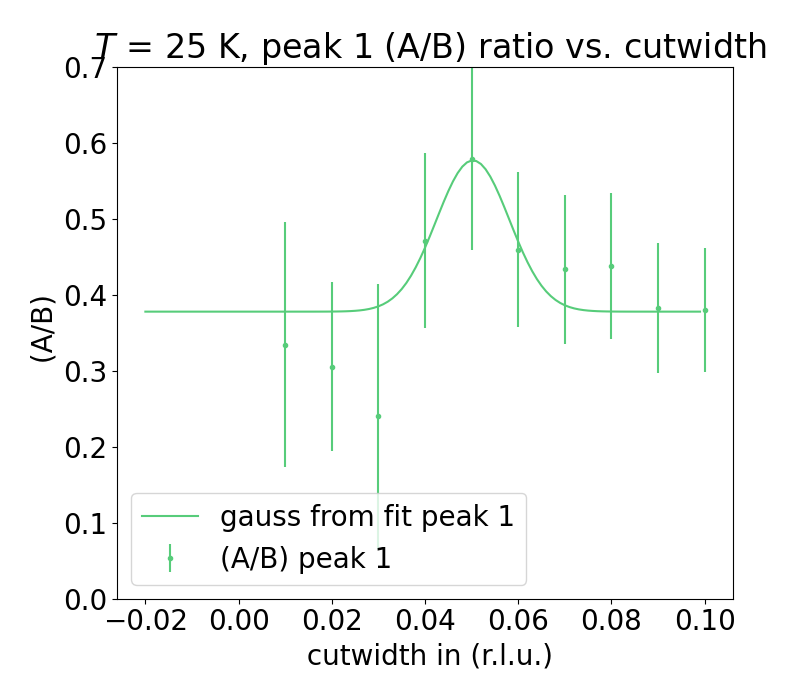}
    \label{A/B 25K peak 1}}
    \hfill
    \subfloat[]{\includegraphics[width=0.45\textwidth]{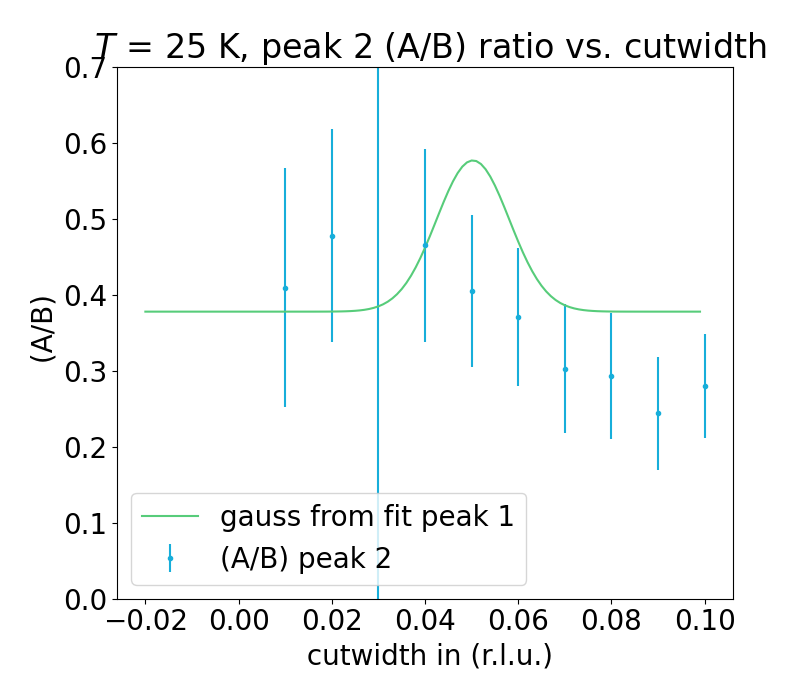}
    \label{A/B 25K peak 2}}

    \subfloat[]{\includegraphics[width=0.45\textwidth]{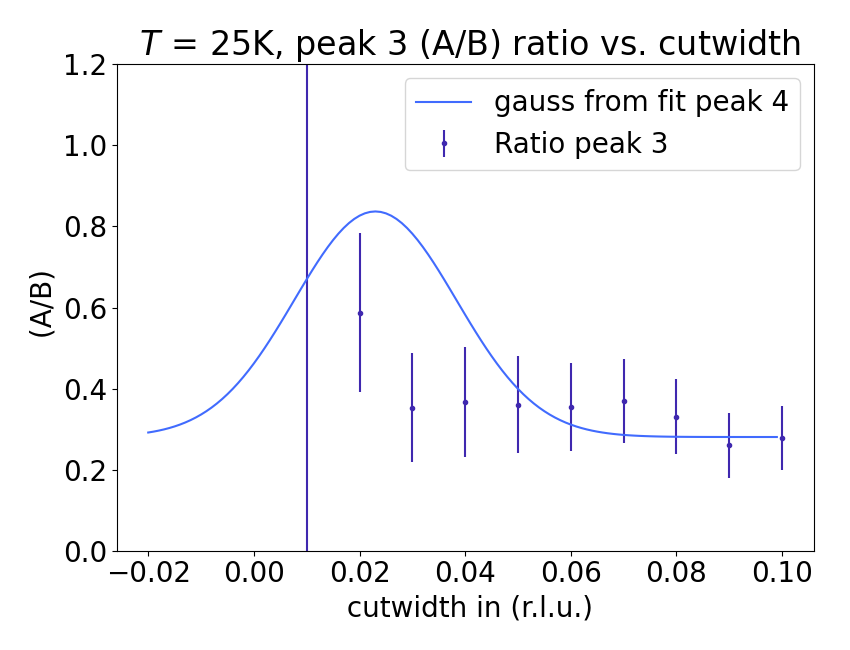}
    \label{A/B 25K peak 3}}
    \hfill
    \subfloat[]{\includegraphics[width=0.45\textwidth]{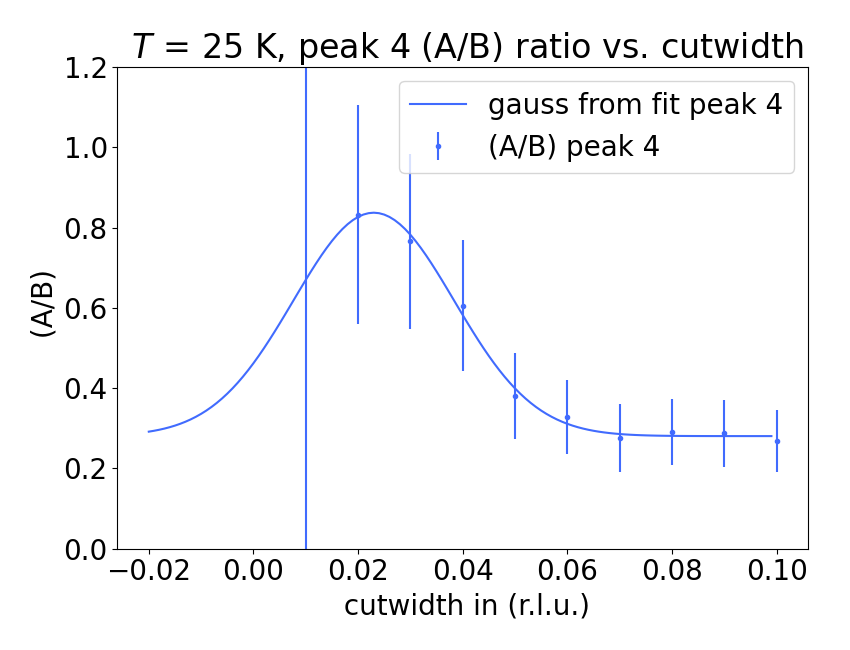}
    \label{A/B 25K peak 4}}
    
    \caption{Determination of the optimal integration width in $q$ for the 25~K data. The amplitude-to-background ratio (A/B) and its dependence on the momentum integration width were examined for peaks 1-4 located at $\Qone$=($1/2 \pm \delta$,1/2,0)  and $\Qthree$=($1/2$,$1/2 \pm \delta$,0) with $\delta\approx0.12$. The optimal integration width was determined to the value where A/B no longer improves. The result was also fitted with a Gaussian lineshape for peaks 1 and 4. The fit was overlayed with the data of peaks 2 and 3 exhibiting  the same optimal integration width in $q$.}
    \label{A/B 25K}
\end{figure}

$Q$-cuts with varying integration widths between 0.01 and 0.1~(r.l.u.) were generated around the four spin-density wave wavevectors,  which we labeled peak 1 to 4 for $\Qone$=($1/2 \pm \delta$,1/2,0)  and $\Qthree$=($1/2$,$1/2 \pm \delta$,0), respectively (as examples of $Q$-cuts see Fig.~\ref{h cut} and Fig.~\ref{k cut}). They were fitted with two Gaussian lineshapes on a constant background. We calculated the ratio between the amplitude and background (A/B) for each peak and plotted them against the integration width (see Fig.~\ref{A/B 40K},~\ref{A/B 30K},~\ref{A/B 25K}). The optimal integration width in $Q$ was chosen to the value where the A/B ratio no longer improves. We also fitted one of the resulting peak pairs of the $H$-cuts and $K$-cuts to a Gaussian on a sloping background. These fits were overlayed with the results of the other two peaks (see Fig. \ref{A/B 40K} for instance). From these results we concluded that identical integration widths for a particular $Q$-direction can be used. The overall results are summarized in Table \ref{table A/B 40K}, showing that an integration width of 0.025~(r.l.u.) is ideal for both $H$ and $K$ cuts at all temperatures.\\

\begin{table} [H]
\begin{center}
    \caption{Chosen energy range at different temperatures to determine the optimal $\bf{Q}$-integration width}
    \label{table T vs E}
	\begin{tabular}{ |c|c|c|c| }
	\hline
	    Temperature & 40~K & 30~K & 25~K \\
	\hline 
	    energy integration width in (meV) & [-1.9, -0.9] & [0.5, 1.5] & [-0.75, -0.15] \\
    \hline
	\end{tabular}
	\end{center}
\end{table}

\begin{table} [H]
\begin{center}
    \caption{The optimal integration width in the $H$ (peaks 3 and 4) or $K$ direction (peaks 1 and 2) was determined for various temperatures by extracting the Gaussian center of the A/B ratio against cut width. The error is the standard deviation of the fit.}
    \label{table A/B 40K}
	\begin{tabular}{ |c|c|c| }
	\hline
	   Temperatures & cut width of peak 1 and 2 & cut width of peak 3 and 4\\
	\hline 
	   40~K & 0.033 $\pm$ 0.007~(r.l.u.) & 0.026 $\pm$ 0.026~(r.l.u.)\\
        \hline 
	   30~K & 0.013 $\pm$ 0.036~(r.l.u.) & 0.023 $\pm$ 0.023~(r.l.u.) \\
        \hline 
	   25~K & 0.05 $\pm$ 0.005~(r.l.u.) & 0.023 $\pm$ 0.015~(r.l.u.) \\
    \hline
	\end{tabular}
	\end{center}
\end{table}

\section*{\texorpdfstring{S\MakeLowercase{upplementary} N\MakeLowercase{ote}}{Supplementary Note} 3. 
C\MakeLowercase{ombination of} $H$- \MakeLowercase{and} $K$-\MakeLowercase{cuts}}
At 40~K, we compared the center and the full-width at half maximum (FWHM) of all $H$, $K$-cuts -- see supplementary  Fig.~\ref{FWHM} and Fig.~\ref{C}. From this, we conclude that the peaks are statistically indistinguishable. On this basis, we have folded the $H$ and $K$ scans into single branches. This operation improves counting statistics while retaining the peak position and peak width. In our analysis, we have applied this operation to all temperatures. 

\begin{figure}[H]
    \centering
    \subfloat[]{\includegraphics[width=0.45\textwidth]{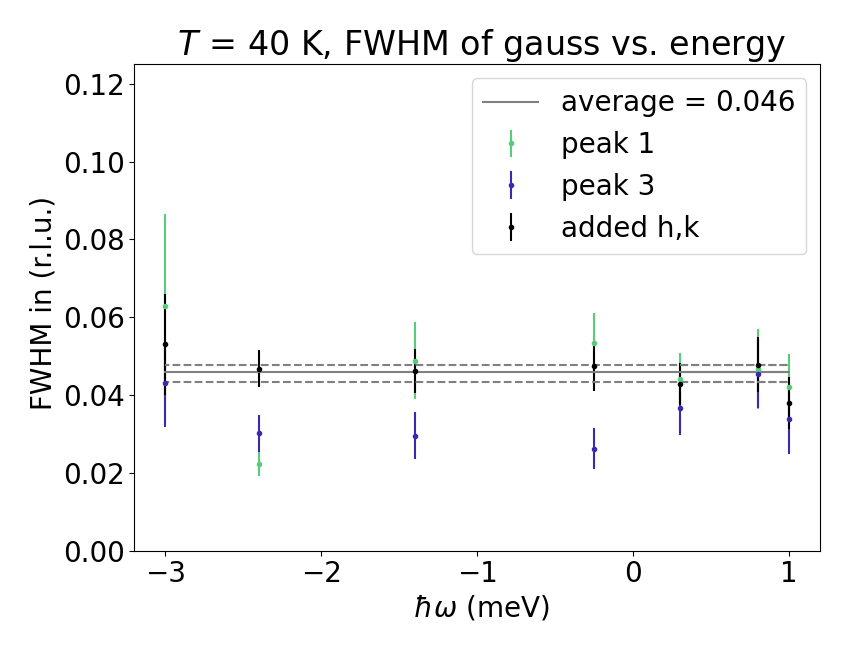}
    \label{FWHM 1}}
    \hfill
    \subfloat[]{\includegraphics[width=0.45\textwidth]{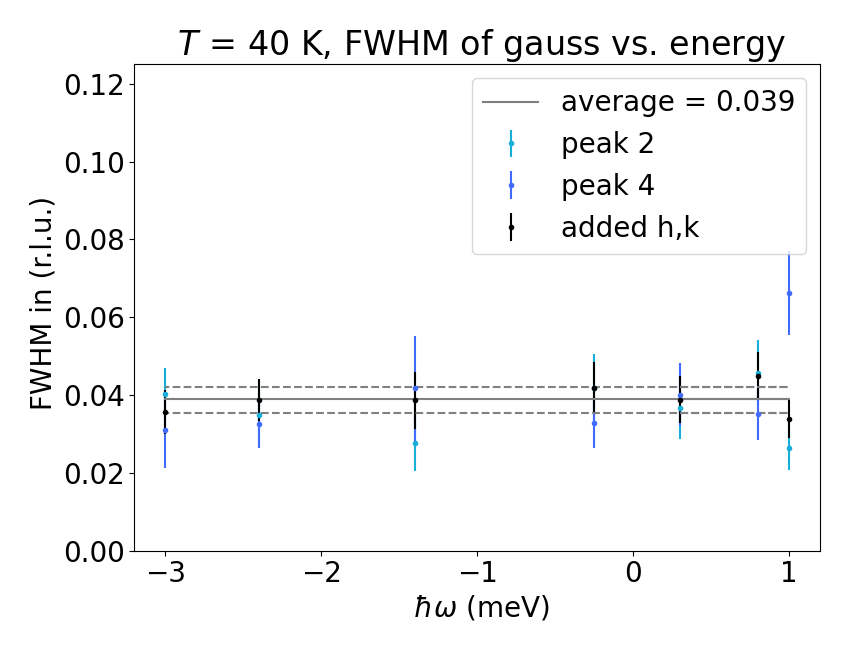}
    \label{FWHM 2}}
    \caption{\textbf{(a)} Comparison of FWHM for peak 1, 3 and their combination. \textbf{(b)} Comparison of FWHM for peak 2, 4 and their combination.}
    \label{FWHM}
\end{figure}

\begin{figure}[H]
    \centering
    \subfloat[]{\includegraphics[width=0.45\textwidth]{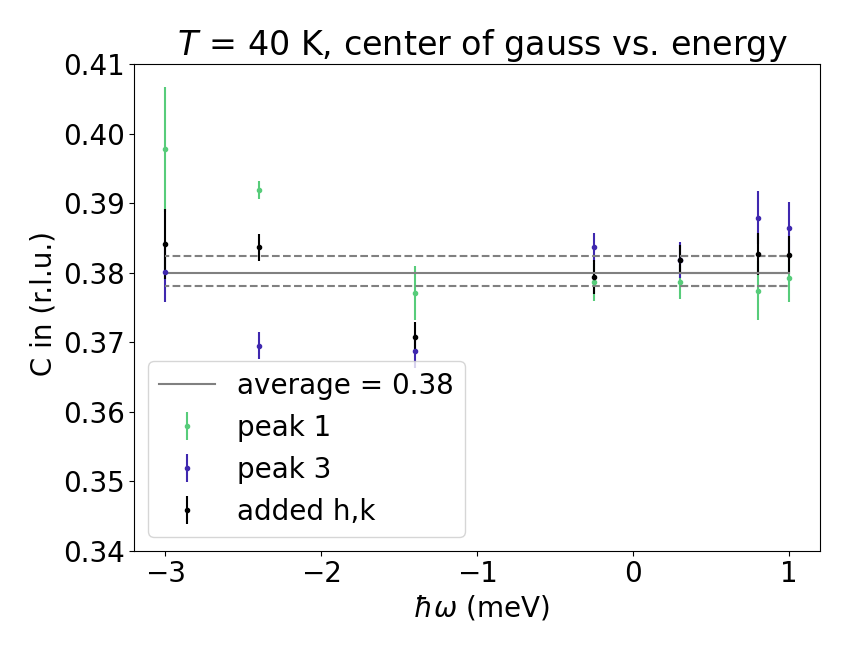}
    \label{C 1}}
    \hfill
    \subfloat[]{\includegraphics[width=0.45\textwidth]{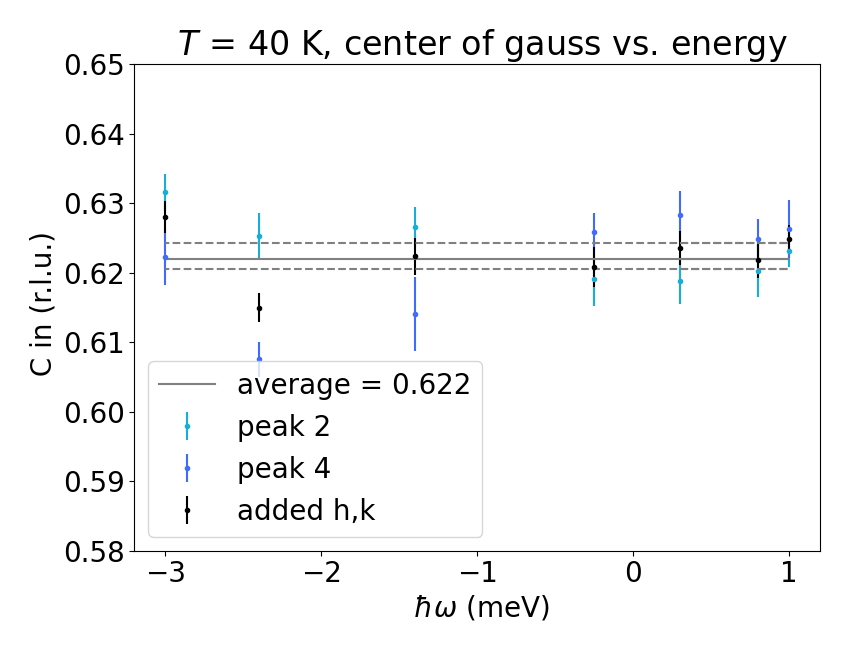}
    \label{C 2}}
    \caption{\textbf{(a)} Comparison of center $C$ for peak 1, 3 and their combination. \textbf{(b)} Comparison of center $C$or peak 2, 4 and their combination.}
    \label{C}
\end{figure}

\section*{\texorpdfstring{S\MakeLowercase{upplementary} N\MakeLowercase{ote}}{Supplementary Note} 4. D\MakeLowercase{etailed} B\MakeLowercase{alance}}
\begin{figure}[H]
 	\begin{center}
 	\includegraphics[width=1\textwidth]{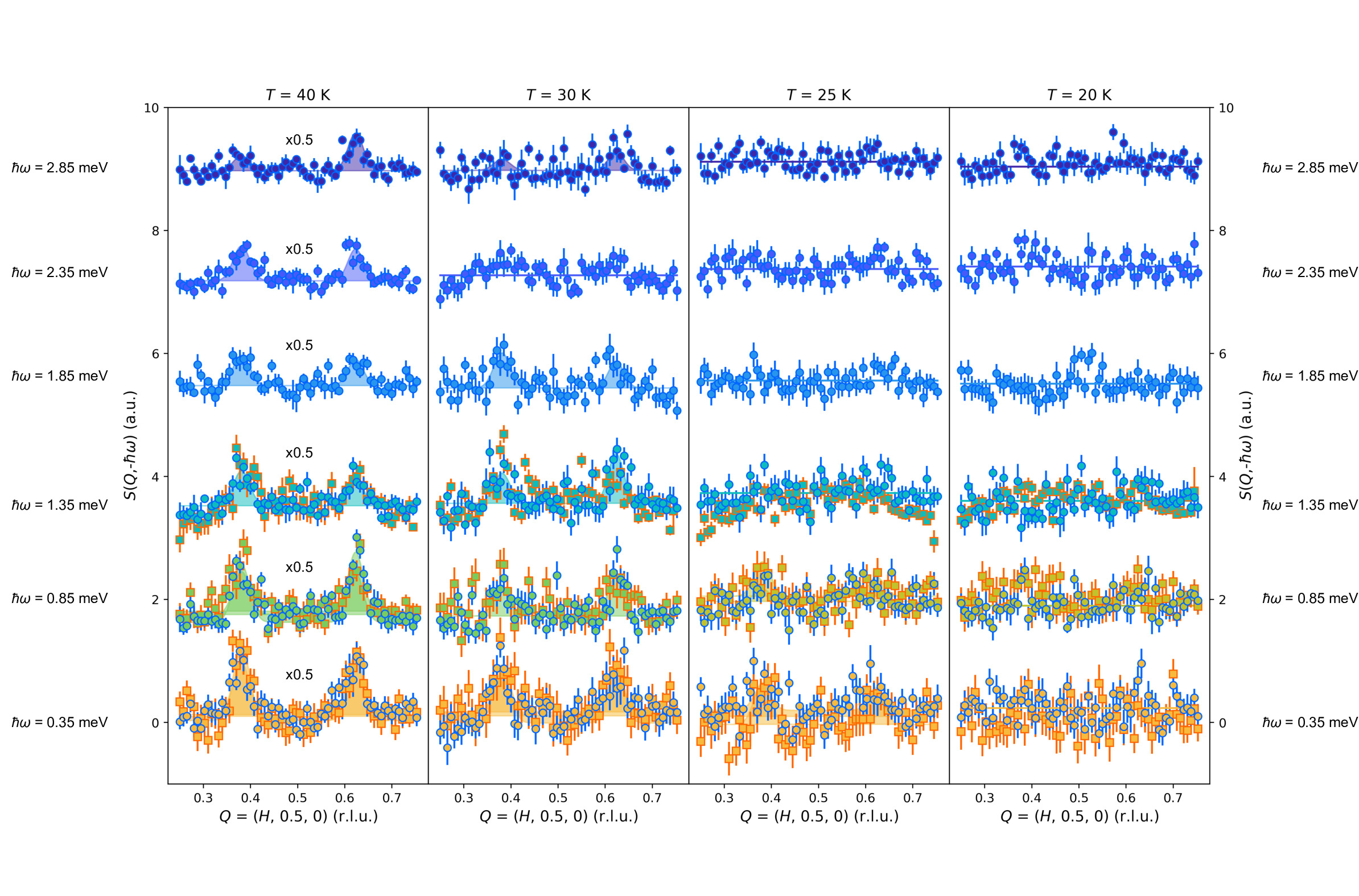}
        \caption{Dynamic structure factor as function of \mbox{$Q$ = ($H$, 0.5, 0) (r.l.u.)} for different energy transfers and temperatures. Here $\bf{Q}$ \mbox{= ($H$, 0.5, 0)} represents a reciprocal space cut through the spin excitations for which the intensity of the $H$ and $K$ cuts were combined. For the sake of visibility, each of the $Q$-cuts have been offset.  Solid lines are Gaussian fits. For the energies \mbox{$\hbar\omega$ = 1.35}, 0.85 and 0.35~meV detailed balance and a constant background subtraction was used to combine energy gain (blue) and loss data (orange) cuts.}
        \label{Fig:Detailedbalance}
 	\end{center}
\end{figure}

An assessment of the $Q$-cuts at negative and positive energy transfers revealed that the data can be further reduced using detailed balance: $S(-\bf{Q},-\hbar\omega)$ = $n_\text{B}(T) S(\bf{Q},\hbar\omega)$ with $n_\text{B}(T)$ = exp(-$\hbar\omega$/$k_\text{B}T$). Figure \ref{Fig:Detailedbalance} shows various $Q$-cuts at different energy transfers and temperatures. For the cuts where a corresponding energy on the energy loss side was measured both cuts are overlayed, $i.e$ the energy gain spectrum and the corrected energy loss cuts. We note that we needed to further correct for a constant background, stemming form background contributions attributed to the sample environment. Notably, detailed balance corrects for the temperature of the sample but not the one of the cryostat usually remaining around 200 K.

\section*{\texorpdfstring{S\MakeLowercase{upplementary} N\MakeLowercase{ote}}{Supplementary Note} 5. A\MakeLowercase{mplitude vs. }E\MakeLowercase{nergy} P\MakeLowercase{lots} }
The spectral weight of the spin excitations was investigated at the fitted $Q$-positions of the combined $H$- and $K$ (see Supplementary Note 3) Figure~\ref{I vs E} shows the spectral weight at $T$ = 40, 30, 25 and 20~K as a function of energy using  an energy step size of 0.1 meV, energy integration width of 0.1~meV in the range of \mbox{$E$ = -3.15} to \mbox{$E$ = -0.15~meV} and \mbox{$E$ = 0.15} to \mbox{$E$ = 1.65~meV} (omitting the elastic line). $Q$ was integrated over \mbox{$Q$ = [0.4875, 0.5125]}~rlu. The individual points of these figures result from a two step fitting process. Firstly, individual $Q$-cuts were fitted with two Gaussians. The weighted average of the resulting FWHM and center $C$ of the 40~K data were then used as fixed parameters to refine the amplitudes, background and slope of the two Gaussians for the data at all temperatures. Figure \ref{I vs E} shows the obtained amplitudes as function of energy at different temperatures. Since the spectral weight of both reduced peaks match, we concluded that the spectral weight of all four excitation branches can be folded onto a single branch.

\begin{figure}[H]
    \subfloat[]{\includegraphics[width=0.45\textwidth]{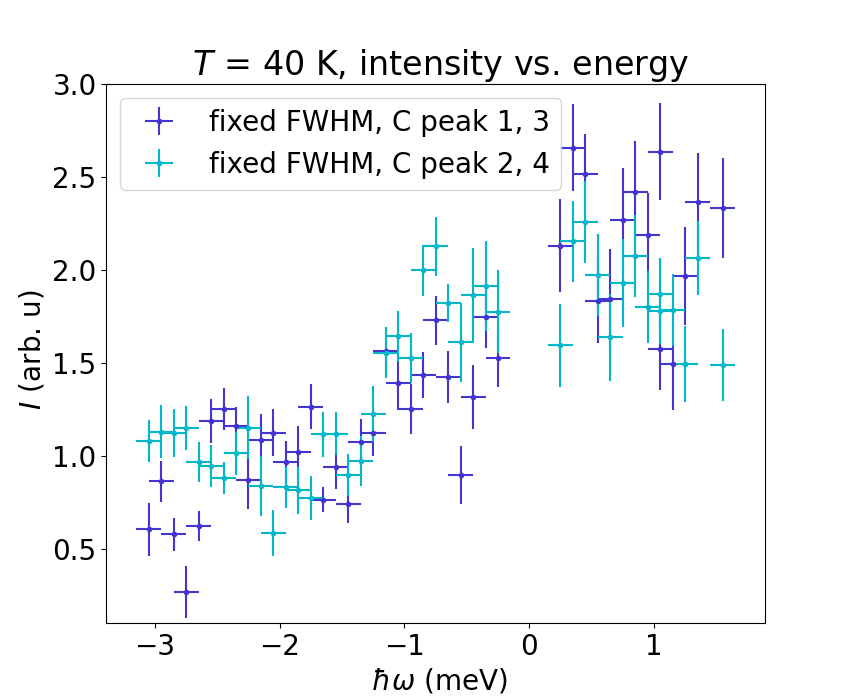}}
    \label{I vs E 40K}
    \hfill
    \subfloat[]{\includegraphics[width=0.45\textwidth]{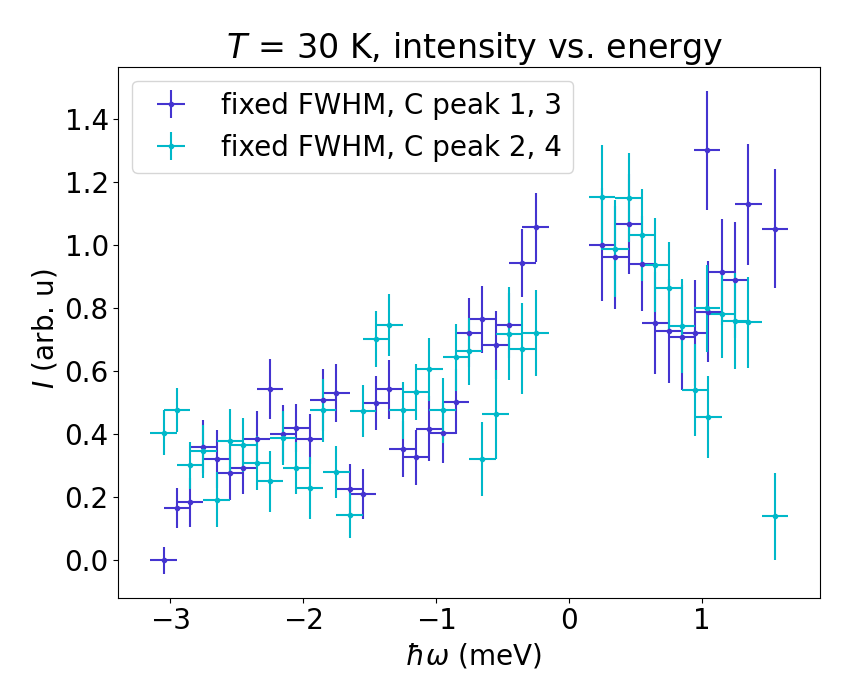}}
    \label{I vs E 30K}

    \subfloat[]{\includegraphics[width=0.45\textwidth]{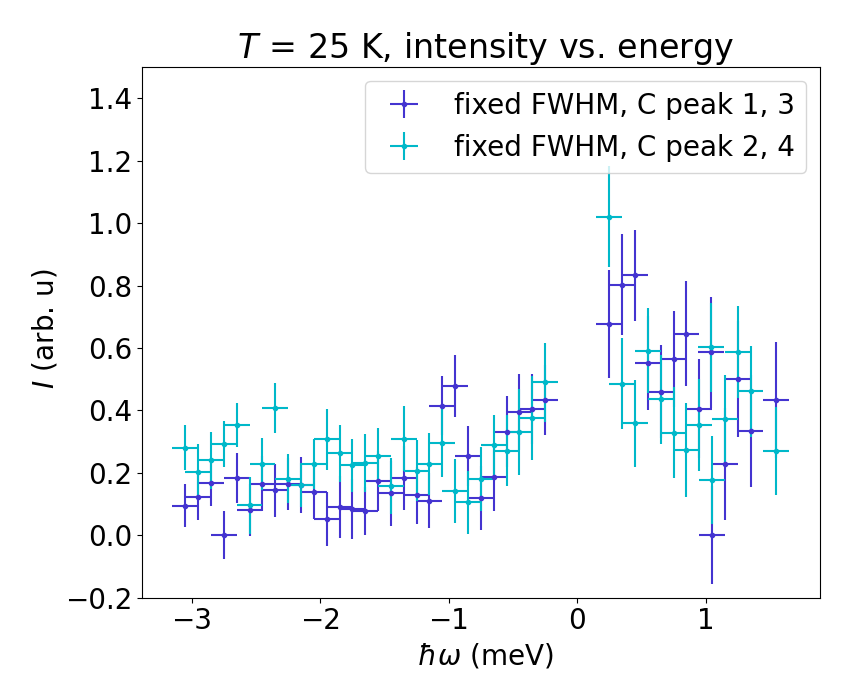}}
    \label{I vs E 25K}
    \hfill
    \subfloat[]{\includegraphics[width=0.45\textwidth]{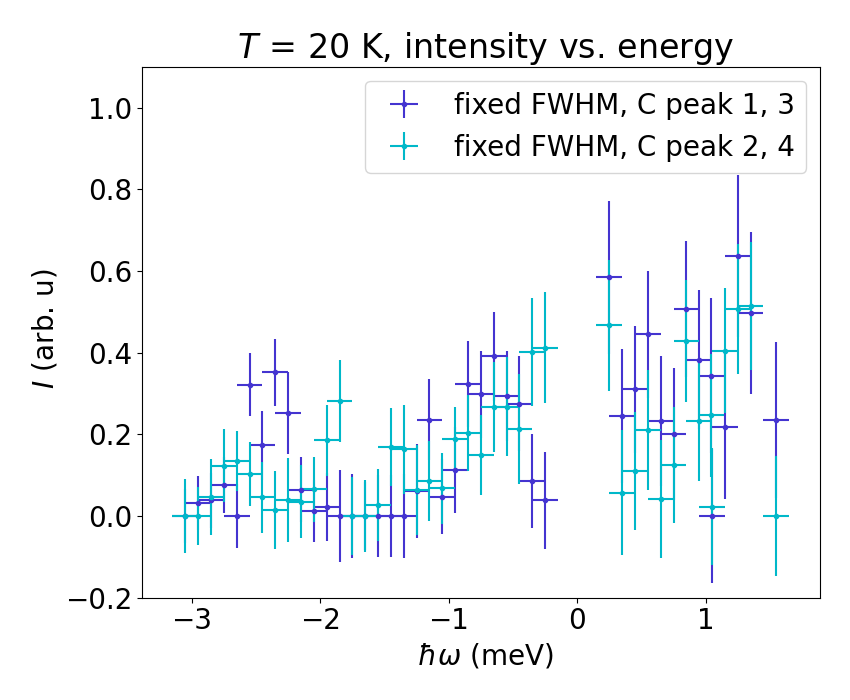}}
    \label{I vs E 20K}
    
    \caption{Spectral weight of LSCO at various temperatures as function of energy for peak 1,3 and 2,4.  The error bar in $x$ direction shows the integration width of the energy. It shows that the spectral weight of all four excitation branches can be mapped onto a single branch.}
    \label{I vs E}
\end{figure}

\section*{\texorpdfstring{S\MakeLowercase{upplementary} N\MakeLowercase{ote}}{Supplementary Note} 6. 
T\MakeLowercase{emperature dependent suppression of low-energy spectral weight}}

\begin{figure}[H]
 	\begin{center}
 	\includegraphics[width=0.9\textwidth]{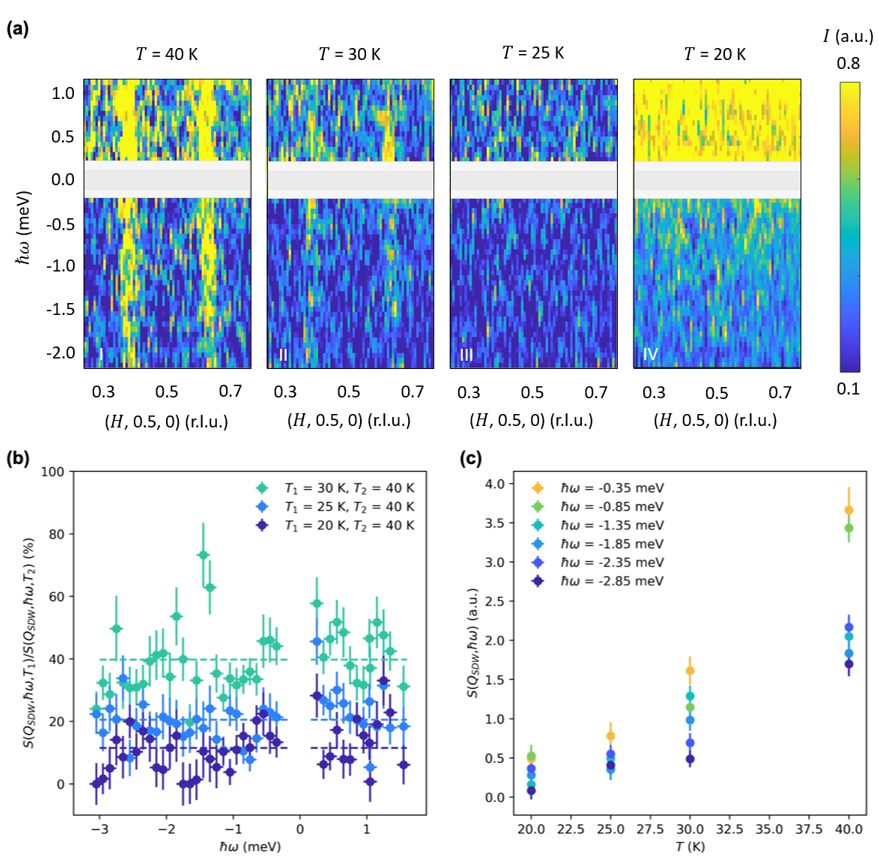}
        \caption{\textbf{(a)} Bose-corrected intensity difference maps along the ($H$, 0.5, 0) and energy-transfer axes, resulting from Fig. 2 of the main manuscript. The Bose-corrected $T$ = 40, 30 and 25 K datasets have been subtracted by the $T$ = 20 K dataset. The  $T$ = 20 K dataset is shown for reference. \textbf{(b)} Relative suppression of the spectral weight at $S$($Q_{SDW}$,$\hbar\omega$) as a function of energy transfer. Here the energy dependence at $T_1$ = 30, 25 and 20 K shown in Fig. 4 of the main manuscript has been normalized by the $T_2$ = 40 K dataset. The dashed lines highlight the statistical mean of the relative spectral weight strengh. \textbf{(c)} Temperature dependence of the spectral weight $S$($Q_{SDW}$,$\hbar\omega$) at $\hbar\omega$  = -0.35, -0.85, -1.35, -1.85, -2.35 and -2.85 meV. }
        \label{Fig:supression}
 	\end{center}
\end{figure}

In this section we add supplementary material for the temperature dependent suppression of the low-energy spectral weight. Figure \ref{Fig:supression}a  shows a temperature difference plot of the neutron intensity along the ($H$, 0.5, 0) and energy-transfer axes. The plot results from the data shown in Fig. 2 of the main manuscript, for which the Bose-corrected $T$ = 40, 30 and 25 K datasets were subtracted with the Bose-corrected $T$ = 20 K dataset.
In panel (b) we report on the relative suppression of the spectral weight at $S$($Q_{SDW}$,$\hbar\omega$) as a function of energy transfer. Here the energy dependence at $T_1$ = 30, 25 and 20 K shown in Fig. 4 of the main manuscript has been normalized by the $T_2$ = 40 K dataset. Within the employed counting statistics we find an equivalent suppression of the low-energy spectral weight across all energy transfers. Notably, the mean spectral weight is suppressed by 60, 80 and 90 \% when the $T$ = 35, 25 and 20 K data are compared to the $T$ = 40 K data. In panel (c) we show the temperature dependence of $S$($Q_{SDW}$,$\hbar\omega$) at specific energy transfers, corroborating the finding of panel (b). We note that the chosen $\hbar\omega$ match the ones of the Q-cuts in Fig. 3 of the main manuscript.




\putbib[ref_SI]
\end{bibunit}
\end{document}